\documentclass[fleqn,usenatbib]{mnras}

\usepackage{newtxtext,newtxmath}

\usepackage[T1]{fontenc}

\DeclareRobustCommand{\VAN}[3]{#2}
\let\VANthebibliography\thebibliography
\def\thebibliography{\DeclareRobustCommand{\VAN}[3]{##3}\VANthebibliography}

\usepackage[usenames,dvipsnames]{color}
\usepackage{graphicx}	
\usepackage{amsmath}	
\usepackage{enumerate}
\usepackage{enumitem}
\usepackage{siunitx}
\usepackage{CJK}

\newcommand{\Sref}[1]{Section \ref{#1}}
\newcommand{\Tref}[1]{Table \ref{#1}}

\sfcode`\.=1001\sfcode`\?=1001\sfcode`\!=1001
\newcommand{\Fref}[1]{\ifhmode \ifnum\spacefactor=1001 Figure \ref{#1}\else Fig.\ \ref{#1}\fi \else Figure \ref{#1}\fi}
\newcommand{\Eref}[1]{\ifhmode \ifnum\spacefactor=1001 Equation (\ref{#1})\else equation (\ref{#1})\fi \else Equation (\ref{#1})\fi}
\newcommand{\Teffnom}{\TextOrMath{$T^_{\mathrm{eff}\odot}$}{T_{\mathrm{eff}\odot}}}
\newcommand{\Teff}{\ensuremath{T_{\mathrm{eff}}}}
\newcommand{\ALi}{\ensuremath{A(\mathrm{Li)}}}
\newcommand{\EPIC}{\textsc{epic}}
\DeclareSIUnit\angstrom{\mbox{\normalfont\AA}}
\newcommand{\besancon}{Besan\c{c}on}

\title[SDST III: new solar twins]{Survey for Distant Solar Twins (SDST) -- III. Identification of new solar twin and solar analogue stars}

\author[C. Lehmann et al.]{Christian Lehmann,$^{1}$\thanks{E-mail: clehmann@swin.edu.au}
Michael T. Murphy,$^{1}$
Fan Liu (刘凡),$^{1,2,3}$
Chris Flynn,$^{1,4}$
Daniel Smith$^{1,5}$ \newauthor
and Daniel A. Berke$^{1,6}$
\\
$^{1}$Centre for Astrophysics and Supercomputing, Swinburne University of Technology, Hawthorn, Victoria 3122, Australia 
\\
$^{2}$School of Physics and Astronomy, Monash University, Melbourne, VIC 3800, Australia
\\
$^{3}$ARC Centre for All Sky Astrophysics in 3D, Canberra, ACT 0200, Australia
\\
$^{4}$OzGrav: ARC Centre of Excellence for Gravitational Wave Discovery, Hawthorn, VIC 3122, Australia
\\
$^{5}$Optical Sciences Centre, Swinburne University of Technology, Hawthorn, Victoria 3122, Australia
\\
$^{6}$User Support Division, Gemini Observatory, 670 N A'ohoku Pl, Hilo, HI 96720-2700, USA
}

\date{Accepted XXX. Received YYY; in original form ZZZ}

\pubyear{2023}

\begin{document}
\begin{CJK*}{UTF8}{gbsn}
\label{firstpage}
\pagerange{\pageref{firstpage}--\pageref{lastpage}}
\maketitle

\begin{abstract}
The Survey for Distant Solar Twins (SDST) aims to find stars very similar to the Sun at distances 1--$\SI{4}{kpc}$, several times more distant than any currently known solar twins and analogues. The goal is to identify the best stars with which to test whether the fine-structure constant, $\alpha$, varies with dark matter density in our Galaxy. Here we use \EPIC, our line-by-line differential technique, to measure the stellar parameters -- effective temperature $\Teff$, surface gravity $\log g$, metallicity [Fe/H] -- from moderate resolution ($R\lesssim32{,}000$) spectra of 877 solar twin and analogue candidates (547 at 1--$\SI{4}{kpc}$) observed with the HERMES spectrograph on the Anglo-Australian Telescope. These are consistent with expectations for $\Teff$ and $\log g$ from photometry, and for [Fe/H] from the \besancon\ stellar population model. \EPIC\ provides small enough uncertainties ($\sim\SI{90}{\kelvin}$, $\SI{0.08}{dex}$, $\SI{0.05}{dex}$, respectively), even at the low signal-to-noise ratios available ($S/N\ga$25\,per pixel), to identify 299 new solar analogues ($\ge90\%$ confidence), and 20 solar twins ($\ge$50\% confidence), 206 and 12 of which are at 1--$\SI{4}{kpc}$. By extending \EPIC\ to measure line broadening and lithium abundance from HERMES spectra, and with ages derived from isochrone fitting with our stellar parameters, we identify 174 solar analogues at 1--$\SI{4}{kpc}$ which are relatively inactive, slowly rotating, and with no evidence of spectroscopic binarity. These are the preferred targets for follow-up spectroscopy to measure $\alpha$.
\end{abstract}

\begin{keywords}
instrumentation: spectrographs -- methods: data analysis -- stars: solar-type -- stars: fundamental parameters -- techniques: spectroscopic
\end{keywords}



\section{Introduction}
The aim of the Survey for Distant Solar Twins (SDST) is to find and spectroscopically confirm distant (up to $\SI{4}{kpc}$) solar twins and analogues. This paper (the third in a series) presents the new solar twin and analogue stars from our observations with the High Efficiency and Resolution Multi-Element Spectrograph (HERMES) at the $\SI{3.9}{\metre}$ Anglo-Australian Telescope (AAT).
The survey design, observations, data processing and verification of survey data products are presented in \citet[][hereafter SDST2]{Liu2022}.
The \EPIC\ algorithm to determine stellar parameters differentially using a solar reference spectrum is discussed in \citet[][hereafter SDST1]{CLehmann2022}. This method is capable of determining stellar parameters with low uncertainties (e.g.\ $\sigma_{\Teff}<\SI{100}{\kelvin}$) for target stars using low signal-to-noise ratio ($S/N\gtrsim20$ in the red CCD) HERMES spectra with medium resolving power ($21{,}000\lesssim R \lesssim 32{,}000$).

A major goal of SDST is to identify solar twins and analogues as probes for variations of electromagnetism's strength, the fine-structure constant $\alpha$, across our Galaxy. The method to determine $\alpha$ is described in \citet[][]{Berke2022b, Berke2022a} and \citet{Murphy2022} and uses precise measurements of the wavelength separation of absorption line pairs in stars. 
Solar analogues are particularly useful for these applications because they have many suitable absorption line pairs. For this purpose we select pairs of deep, unsaturated absorption lines with small separations ($<\SI{800}{\kilo\metre\per\second}$), and known sensitivity to variations in $\alpha$ \citep{Dzuba2022}. Their separations can also be compared with a multitude of high S/N local solar analogue and Sun spectra to measure any changes in $\alpha$ between stars \citep[][]{Berke2022b, Berke2022a}.
We adopt the following definitions of solar twins and analogues from \citet[][]{Berke2022b}:
\begin{align}
    \text{Solar twin} &= \left\{
    \begin{array}{r@{\,}l}\label{eq:ST_def}
         \Teffnom&\pm\,\SI{100}{\kelvin}, \\
         \log{g_\odot}&\pm\,\SI{0.2}{dex},\\
         \mathrm{[Fe/H]_\odot}&\pm\,\SI{0.1}{dex},\\
    \end{array} \right.
\end{align}
\begin{align}
    \text{Solar analogue} &= \left\{
    \begin{array}{r@{\,}l}\label{eq:SA_def}
        \Teffnom&\pm\,\SI{300}{\kelvin},\\
        \log{g_\odot}&\pm\,\SI{0.4}{dex},\\
        \mathrm{[Fe/H]_\odot}&\pm\,\SI{0.3}{dex}\\
    \end{array} \right.
\end{align}

We have observed 877 solar twin candidate stars in a direction close to the Galactic centre.
We present two major results in this work: (i) the target stars with the highest probability to be solar twin and analogue stars and (ii) the stars which are best suited for observations of the fine-structure constant $\alpha$. The main difference between these groups is that active stars, or stars with much broader lines than the Sun (from, e.g., rotation or spectroscopic binarity), are not suited to be probes for $\alpha$ while they can still be solar twin candidates.

This paper is structured in the following way.
In \Sref{sec:SDST2_sum} we summarise the survey observations and basic characteristics such as the $S/N$ distribution.
\Sref{sec:activity_indicator} presents changes to the $\EPIC$ algorithm to measure additional parameters such as the lithium abundance and the spectral line broadening.
In \Sref{sec:results} we present the results of the survey, including the stellar parameters and additional activity indicators, and also the catalogue which includes 299 new solar analogue stars. 
\Sref{sec:conclusion} summarises the main results of this work and explains our future plans with these solar analogue stars.

\section{SDST observations with HERMES}\label{sec:SDST2_sum}
\begin{figure}
    \centering
    \includegraphics[width=0.445\textwidth]{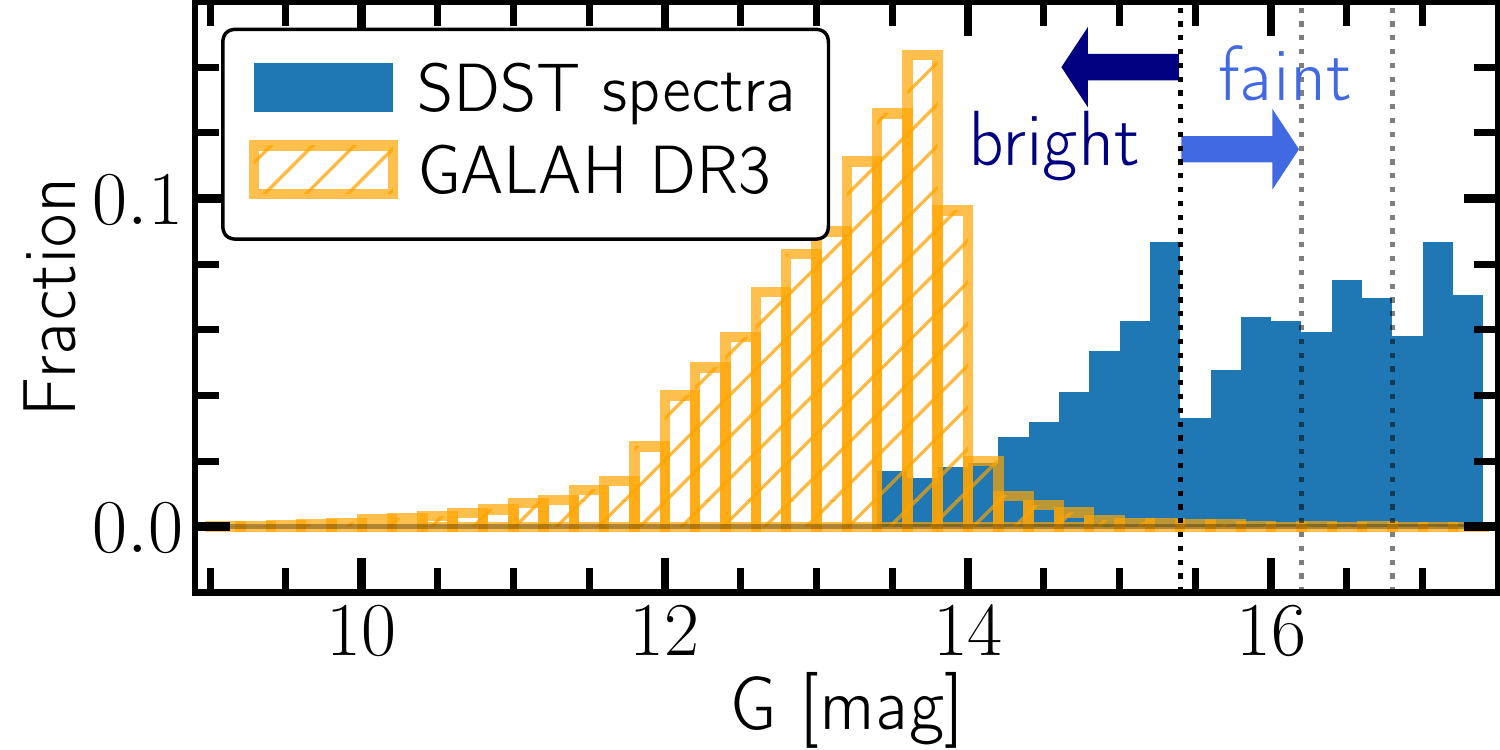}
    \caption{Normalised $G$ magnitude distribution of the SDST sample (877 stars; filled blue bars) and solar analogue candidates in GALAH DR3 (unfilled orange bars). Our sample is separated into 330 bright ($G\leq15.4$) and 547 faint target stars ($G>15.4$). The faint stars further split into three $G$ magnitude bins (split at $G=16.2$ and $G=16.8$) which were used to obtain fairly uniform $S/N$ (SDST2), i.e.\ more exposures were allocated to fainter targets.}
    \label{fig:Gmag_hist}
\end{figure}
This section summarises information about the observations at the HERMES/AAT facility, which can be found in detail in SDST2.
Within this survey, we observe a 2$^\circ$-field off the Galactic Disk and adjacent to the Galactic Centre, i.e.\ Galactic longitude $329.13^\circ<b<330.86^\circ$ and Galactic latitude $15.68^\circ<l<17.43^\circ$, with the HERMES spectrograph \citep[][]{Sheinis2015} on the $\SI{3.9}{\metre}$ AAT. This field was chosen to probe further towards the Galactic Centre, which contains a higher dark matter density \citep[][]{Sofue2012, Ablimit2020} than the solar neighbourhood so that we can test for a connection between it and variations in the fine-structure constant while also avoiding dust/extinction \citep[][]{Sahlholdt2021}.
We observed 330 bright ($\SI{13.4}{mag}\leq G< \SI{15.4}{mag}$) and 547 faint ($\SI{15.4}{mag}\leq G< \SI{17.4}{mag}$) solar twin and analogue candidates; the full selection process is described in SDST2. 
The fraction of stars observed in each magnitude range by SDST as well as the distribution of likely solar analogue stars in GALAH DR3 \citep{Buder2021}, the main survey making use of the HERMES instrument, are shown in \Fref{fig:Gmag_hist}.

The bright targets were observed exclusively during evening twilight when it was not possible to observe faint targets. Poor weather conditions and instrument set-up delays meant only low $S/N$ spectra were obtained for the bright targets, so we do not consider them as part of our primary results.
The faint targets were observed over 14 half-nights (7 of which provided useful data), and the total number of exposures was adjusted according to their brightness to achieve relatively uniform $S/N$ over the magnitude range of the faint target bins, i.e. the faintest targets ($\SI{16.8}{mag}\leq G< \SI{17.4}{mag}$) were observed in all 23 primary exposures while brighter targets ($\SI{15.4}{mag}\leq G< \SI{16.2}{mag}$) were observed in only 5--8 exposures.

The main challenge for the SDST observations was to build up $S/N\gtrsim20$ per pixel in the $R$ (red) band. For this reason, and to simplify the analysis with $\EPIC$, we combined multiple exposures of each star, in each band. This enables us to achieve the target $S/N$ in 88\% of faint target stars. We discuss the resulting $S/N$ and uncertainties in \Sref{sec_SP_uncertainties}.

\subsection{Resolving power for combined spectra}\label{sec:resolving_power}
The resolving power of each spectrum is needed for accurate comparison with the solar reference spectrum when measuring equivalent widths of absorption lines with $\EPIC$ (SDST1, section 2.3.3). The differences in resolving power from fibre to fibre in HERMES are sufficiently large to cause systematic differences in EW measurements, which results in systematic differences in stellar parameter measurements unless carefully corrected. Therefore, it is important to accurately determine the resolving power for each combined spectrum in SDST.

Within the \EPIC\ algorithm, the fibre identification is used to determine the resolving power for each absorption feature within a spectrum according to the resolving power maps of \citet{Kos2017}, as shown in SDST1 (section 2.3.3). When SDST observations from different fibres are combined into a single spectrum, their new wavelength-dependent resolving powers will be within the range of resolving powers of all fibres involved in observing the target. The resulting resolving power is estimated by taking the weighted mean, weighted by the $S/N$ in the fibres, of the resolving power from all contributing spectra.
We tested this approach by simulating spectra with differing resolving power within the range found for HERMES fibres. The resolving power of the combined absorption line was then measured by fitting a Gaussian profile. This differed negligibly from the weighted mean resolving power, with the largest difference being $\sim3\%$ for the most extreme case, i.e.\ an equal weight of $R=21{,}000$ and $R=32{,}000$, which are the lowest and highest resolving powers within HERMES respectively.

\subsection[Estimating the S/N]{Estimating the $S/N$}
We use the $S/N$ as the main indicator for the quality of spectra. The $S/N$ is estimated separately in each spectral band, avoiding both hydrogen features and telluric absorption, and following the same procedure as in SDST2 using the following wavelength ranges:
\begin{align}
    &B\textrm{ band: }  \,\,4730 - \SI{4840}{\angstrom}\\
    &G\textrm{ band: }  \,\,5670 - \SI{5860}{\angstrom}\\
    &R\textrm{ band: }  \,\,6500 - \SI{6540}{\angstrom}, 6590 - \SI{6720}{\angstrom}\\
    &I\textrm{ band: } 7670 - \SI{7870}{\angstrom}\label{eq:SNR_range}.
\end{align}
Within these wavelength ranges, we determine the upper 50\% flux values and their flux uncertainties. The $S/N$ is the median of these flux values (the "75th percentile") divided by the median of their uncertainties.
The 75th percentile was chosen because it avoids both absorption features as well as cosmic rays, so it estimates the $S/N$ of the continuum.

\section{Stellar activity indicators}\label{sec:activity_indicator}
In this section, we describe how we implemented lithium abundance, line broadening and age measurements as indicators for stellar activity in the $\EPIC$ algorithm. The results of these measurements are discussed in \Sref{sec:age_indicator_measure}.

\subsection{Lithium abundance, \ALi}\label{sec:lithium_abundance}
\begin{figure}
    \centering
    \includegraphics[width=0.445\textwidth]{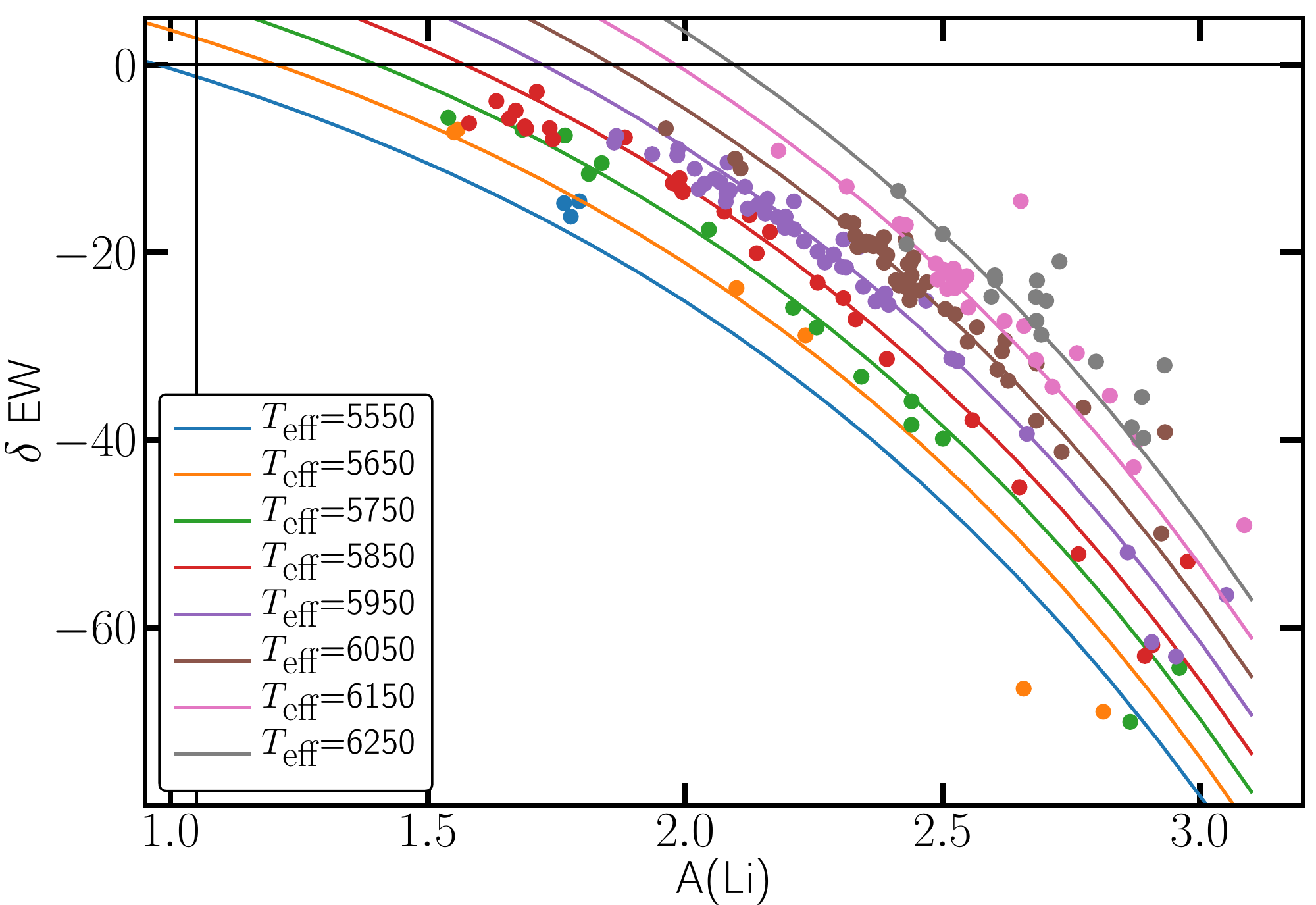}
    \caption{The lithium abundance model. Each point represents a GALAH DR3 spectrum with a corresponding lithium abundance measurement. The coloured lines show the best fits for the deviation from the solar EW, $\delta$EW, of the lithium feature at $\SI{6707}{\angstrom}$ to the lithium abundance [A(Li)] and effective temperature (\Teff). $\delta\textrm{EW} = 0$ means that the feature has the same EW as the solar lithium feature and $\ALi=1.05$ is the solar lithium abundance, both shown as black solid lines. The coloured lines represent model isotherms with $\Teff$ values shown. The $\delta$EW measurements are shown as dots where the colour corresponds to the isotherm with the closest matching temperature, e.g.\ a star with $\Teff = \SI{5567}{\kelvin}$ would be blue. For simplicity, we have selected only stars close to the given temperature of an isotherm ($\pm\SI{25}{\kelvin}$).}
    \label{fig:Li_model}
\end{figure}
The only lithium feature available within the HERMES spectral bands is the doublet at $\SI{6707.81}{\angstrom}$. \EPIC\ measures the EW of this feature and, using measured lithium abundances in GALAH DR3 \citep{Buder2021}, allows us to connect the EWs with \ALi. Note that we have to convert the relative lithium abundance [Li/Fe] in the GALAH DR3 catalogue to an absolute value using 
\begin{align}
    \ALi =  \textrm{[Li/Fe]}  +  \textrm{[Fe/H]}  + 1.05. 
\end{align}
[Li/Fe] is a logarithmic lithium abundance relative to iron, normalised on the solar metal content, so adding [Fe/H] leads to a measurement relative to hydrogen ([Li/H]) and adding 1.05 \citep[the solar lithium abundance, ][]{Lodders2019} estimates an absolute abundance value, i.e.\ \ALi. 
Creating a model of how the EW varies with $\ALi$ is more difficult than for the three main stellar parameters (as done in SDST1) because there is no stellar library on which to apply the \EPIC\ measurement approach. Instead, we use a set of high $S/N$ GALAH spectra. This means that variations from spectrum to spectrum might play a role in the model uncertainties. Within GALAH DR3, there are 4433 spectra that fulfil the following requirements for this model construction step:
\begin{enumerate}[labelwidth=*, leftmargin=*]
    \item The spectrum needs to have a measured [Li/Fe] value in the catalogue. This means that the lithium feature must have been strong enough for a reliable measurement against the noise in each GALAH spectrum, which is $\ALi\gtrsim1.5$ in GALAH DR3 for all Sun-like stars (as defined in iii). This limits reliable $\ALi$ measurements to only high lithium abundance stars.
    \item We select only high $S/N$ spectra to minimise the uncertainty in the model. The cut was made for spectra with $S/N>100$, which is high enough to reliably measure the EW of the lithium feature in the solar spectrum [$\ALi = 1.05$].
    \item The lithium abundance is interesting for us mostly in Sun-like stars, i.e. solar analogues. Therefore, we select spectra with stellar parameters (measured with $\EPIC$) that are rather close to solar values (i.e. $\Teff, \log g$, [Fe/H]$ = 5772\pm\SI{500}{\kelvin}, 4.44\pm\SI{0.5}{dex}, 0.0\pm\SI{0.5}{dex}$).
\end{enumerate}

We found that the lithium feature EW can be modelled with only a dependency on effective temperature $\Teff$ and \ALi:
\begin{align}\label{eq:Li_model}
    \textrm{EW}_\textrm{Li} = a \Teff + b_1 \exp\left\{b_2 \left[A\left(\textrm{Li}\right) - b_3\right]\right\} + c.
\end{align}
\EPIC\ is applied to each of the selected GALAH spectra and fits the $a, b_1, b_2, b_3 \textrm{ and } c$ parameters to the data.
With these parameters, we can measure $\ALi$ through a measurement of the lithium feature EW in a solar-like target star observed with HERMES.
We estimate the lithium abundance uncertainty using \Eref{eq:Li_model} and error propagation from uncertainties in the effective temperature and lithium EW.
We tested how much uncertainties in the model parameters $a, b_1, b_2, b_3 \textrm{ and } c$ contributed in an error propagation and found it to be small ($<1\%$), so, for simplicity, we treat them as negligible.

The model fit with varying temperatures is displayed in \Fref{fig:Li_model}. Each of the points represents one GALAH spectrum in which we measured the lithium feature EW and which satisfies the above criteria. The algorithm attempts to match each of the measurements from the 4433 available spectra. We have tested this model on the spectra it was constructed on (4433 GALAH DR3 spectra) and found their measured lithium abundances to have Gaussian scatter corresponding to model uncertainties around their previous measurements. However, it should be noted that the lowest lithium abundance in this training set is $\ALi\approx1.5$, and we are unable to test any extrapolation of the model below this value. Nevertheless, this is not relevant for our purposes as the goal is to select against high $\ALi$ (potentially young and active) stars.

\subsection{Line broadening}
\begin{figure}
    \centering
    \includegraphics[width=0.445\textwidth]{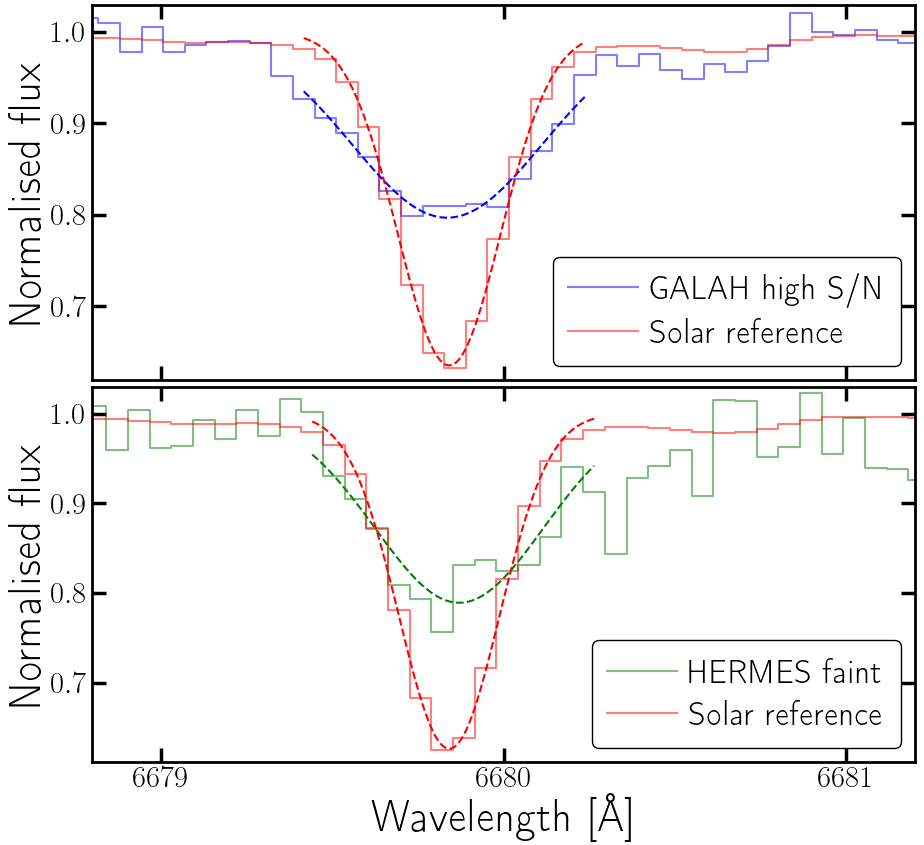}
    \caption{Broadening measurement of two different spectra for an example absorption line. The upper panel shows a spectrum that has a measured broadening of $\SI{12.9}{\kilo\metre\per\second}$. This object (\href{http://simbad.cds.unistra.fr/simbad/sim-id?Ident=UCAC4+257-076579&NbIdent=1&Radius=2&Radius.unit=arcmin&submit=submit+id}{UCAC4 257-076579}) has been confirmed to be a spectroscopic binary, which causes the large observed broadening.
    In the lower panel we show GAIA DR3 6197759819563606016, which is one of our faint sample stars and flagged as broadened ($\SI{8.9}{\kilo\metre\per\second}$). This example demonstrates that a measurement of the broadening from a single line is not reliable in low $S/N$ stars, which is why we use an average broadening. For reference, the measured solar broadening is $\sim\SI{4.9}{\kilo\metre\per\second}$.}
    \label{fig:broadening_ex}
\end{figure}
We determine which lines are most useful for this measurement by initially measuring the line broadening (using the method below) in the solar reference spectrum and selecting against features that appear broadened for other reasons, e.g. because of blending with other features. We find that the average broadening in the Sun is $4.90 \pm \SI{0.01}{\kilo\metre\per\second}$, which is slightly higher than what the literature on solar macroturbulence suggests \citep[$\sim\SI{3.69}{\kilo\metre\per\second}$, ][]{Gray2018}. This is likely due to the medium resolving power spectra ($R\lesssim32{,}000$, FWHM\,$\gtrsim\SI{9.5}{\kilo\metre\per\second}$) of HERMES which limits the line broadening measurements to $\gtrsim\SI{4}{\kilo\metre\per\second}$. However, the purpose of this step is to select against spectra with high line broadening, which means that accurate low broadening measurements are not required.

We measure the broadening of individual features using the inner 13 pixels around the centre of the absorption line (SDST1 section 2.3.2) by fitting the following Gaussian function to these pixels:
\begin{align}
    \mathcal{F} = 1 - a \exp\left[-\frac{(\lambda - \lambda_0)^2}{2(\sigma^2 + \sigma_0^2)}\right],
\end{align}
where $a$ is the line's absorption depth, $\lambda$ is the wavelength centre of each pixel, $\lambda_0$ is the wavelength of the line centre, $\sigma_0$ is the instrumental resolution $(\textrm{FWHM}/2\sqrt{2\ln2})$ and $\sigma$ is the additional broadening. The instrumental resolution of HERMES is known for each absorption line via their resolving power maps \citep{Kos2017}. All $\sigma$ widths are converted into units of $\SI{}{\kilo\metre\per\second}$ for comparison with each other.

We select only absorption lines within our line list (SDST1, section 2.2) which are not blended in the solar reference spectrum, i.e.\ with $\sigma<\SI{5}{\kilo\metre\per\second}$, which leaves up to 89 absorption lines for the broadening measurement. The limit was set by visually inspecting absorption lines for blends and viewing their measured broadening; every blended line was above the $\SI{5}{\kilo\metre\per\second}$ limit. We take the quadrature difference $\sqrt{|\sigma_1^2 - \sigma_2^2|}$ of reference and target spectrum to arrive at a relative difference which can be used to determine how broadened each line is compared to solar features. We then select against outliers in the sample of lines by determining the weighted sum (weighted by the uncertainty of $\sigma$) and the standard deviation of these $\sigma$ differences. Lines deviating more than three standard deviations are eliminated from this sample. The weighted sum is re-computed with the remaining lines to arrive at the final broadening parameter.

\Fref{fig:broadening_ex} shows two example fits to measure the broadening in the same absorption feature in different spectra. The top panel is a GALAH spectrum broadened because the star (\href{http://simbad.cds.unistra.fr/simbad/sim-id?Ident=UCAC4+257-076579&NbIdent=1&Radius=2&Radius.unit=arcmin&submit=submit+id}{UCAC4 257-076579}) is a confirmed spectroscopic binary, where the lines from the two stars are separated by a radial velocity difference. The spectrum has high $S/N$ and therefore we could measure this broadening with only a few absorption lines. The bottom panel shows a star within our distant sample of solar analogue candidates. Its lines are difficult to measure because of low $S/N$, but using many features allows us to determine that this star has, on average, broadened lines, i.e.\ within the upper 5\% of our sample in terms of broadening (see \Sref{sec:age_indicator_measure} for the broadening measurements in the SDST faint stellar sample).

\subsection[Age measurement with q2]{Age measurement with q$^2$}
We make use of stellar parameters measured by $\EPIC$ and the qoyllur-quipu (q$^2$) algorithm \citep{Ramirez2014} to derive stellar ages for SDST targets. q$^2$ uses isochrone fitting with Yonsei-Yale isochrones \citep[][]{Demarque2004} to determine stellar ages.

\section{Results}\label{sec:results}
In this section we present analysis of the fully reduced data-set of 877 target stars for our $2^\circ$-field centred at Galactic longitude $l=330.0^\circ$ and latitude $b=16.6^\circ$. We present spectroscopic stellar parameters (\Teff, $\log g$, [Fe/H]) of these targets, which were calculated with the \EPIC\ algorithm. We verify these stellar parameters by comparing them with photometric \Teff\ and $\log g$ estimates, as well as a [Fe/H] distribution simulated with the \besancon\ model \citep[][]{Robin2012, Czekaj2014, Bienayme2018}. We analyse the stellar parameter uncertainties from $\EPIC$ to give a better understanding of the advantages and limitations of this survey.
Additionally, we present the lithium abundance $\ALi$ and the line broadening, which are investigated together with the ages calculated with the q$^2$ isochrone fitting algorithm \citep{Ramirez2014}. This is relevant to select against potentially active or fast rotating solar twin candidates that would not serve as good targets to measure the fine-structure constant \citep[][]{Berke2022b, Berke2022a}. Lastly, we determine biases that arise as a result of candidate stars being observed at different distances within our field. This is important for assessing how well the SDST performed for the most distant stars probed, and it is informative for future surveys of this kind. The full catalogue for our targets with all the information derived in this work can be accessed in the supporting information online and \citet[][]{CLehmann942022a}.

For all stellar parameter measurements, we present the results for \EPIC\ in two different calibrations: (i) One based on stellar parameters from \citet[][]{Casali2020}, who derived spectroscopic stellar parameters for stars observed at high resolution ($R\approx115{,}000$). These are considered spectroscopic values, which may not necessarily agree with values measured photometrically. (ii) Another based on the library of stacked GALAH (DR2) spectra from \citet[][]{Zwitter2018} whose stellar parameters are calibrated to match photometrically determined surface gravities.

\subsection{New solar twins and analogues}\label{sec:new_ST}
\begin{figure}
    \centering
    \includegraphics[width=0.445\textwidth]{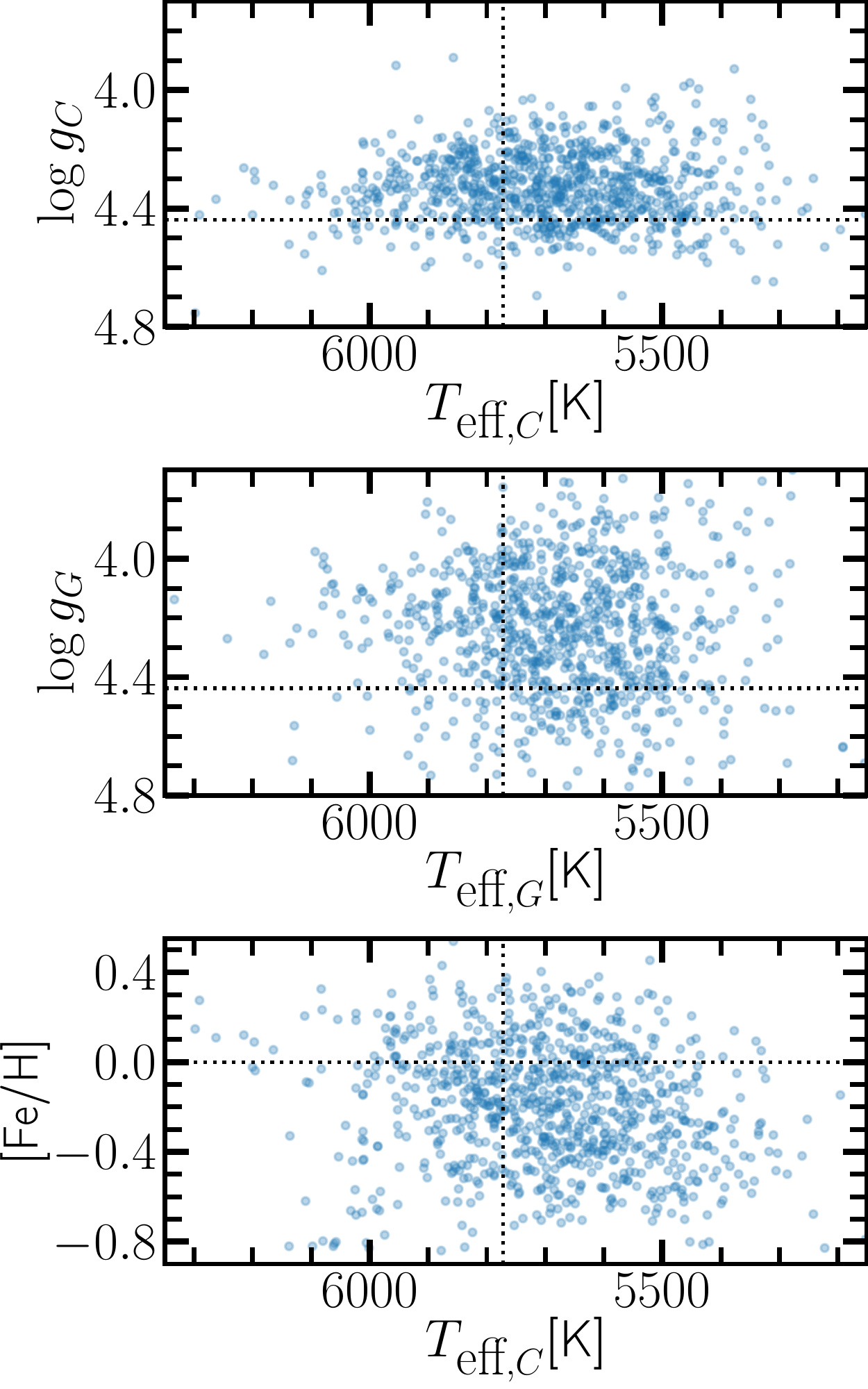} 
    \caption{Effective temperature, surface gravity and metallicity of the 877 SDST sample. These stellar parameters have been computed with the \EPIC\ algorithm and are used to identify how solar-like a target star is. Solar stellar parameters are shown as doted black lines on each panel. The top panel contains $\Teff$ and $\log g$ calibrated with spectroscopic values \citep[][]{Casali2020} while the middle panel is calibrated on GALAH DR2 \citep[][]{Buder2018}. The bottom panel contains the metallicity which is virtually unaffected by re-calibrations.}
    \label{fig:sp_hr}
\end{figure}
In \Fref{fig:sp_hr} we show the stellar parameters calculated for all 877 stars and both calibrations. The spread around solar values in both effective temperature and surface gravity suggests that the photometric pre-selection (SDST2) of targets worked as intended, i.e.\ 87\% of stars are within the solar analogue definition for effective temperature [\Eref{eq:SA_def}], 99\% for surface gravity, and 61\% for metallicity. 
Only 8 stars are potentially red giant or turn-off stars ($\log g_C < 4.0$).

The metallicity values we found for our stars average at $\SI{-0.12}{dex}$ with a large dispersion of $\SI{0.26}{dex}$, and can be seen in the bottom panel of \Fref{fig:sp_hr}. The median of the metallicity distribution in the Solar neighbourhood is between $-0.1$ and $\SI{0.0}{dex}$ \citep[][]{Haywood2001}. Looking 2--$\SI{4}{kpc}$ toward the inner Galaxy, we expect the mean metallicity to rise due to the metallicity gradient in the Galactic disk \citep[e.g.][]{Minchev2018, Ness2019}, but we also expect it to decline because of the Galactic latitude chosen ($15.68^\circ<l<17.43^\circ$) to avoid excessive extinction. Consequently, there are opposing effects on the metallicity distribution. To investigate this in detail, we compare our metallicity distribution with the \besancon\ stellar population model in \Sref{sec:verification} to test our measurements. 
The large spread of metallicities from $-0.8$ to $\SI{0.4}{dex}$ is expected, as metallicity values are difficult to constrain using photometric selections. The metallicity constraint in SDST2 is largely done via the SkyMapper $(v-g)_0$ colours \citep[][Figure 7]{Casagrande2019a}, which is dependent on both $\Teff$ and [Fe/H].

As expected, we find that different calibrations of the $\EPIC$ algorithm yield different results, i.e.\ the initial photometric calibration (on GALAH DR2) is shifted, from a mean of $\langle \log g_C \rangle = \SI{4.33}{dex}$ to $\langle \log g_G \rangle = \SI{4.24}{dex}$, and has a higher dispersion (\SI{0.12}{dex} versus \SI{0.22}{dex}) compared to the spectroscopic calibration with \citet[][]{Casali2020} surface gravity values. This means that the GALAH DR2 calibration will have a larger spread of values and higher uncertainties, which is reflected in lower solar analogue probabilities below.

\begin{figure}
    \centering
    \includegraphics[width=0.445\textwidth]{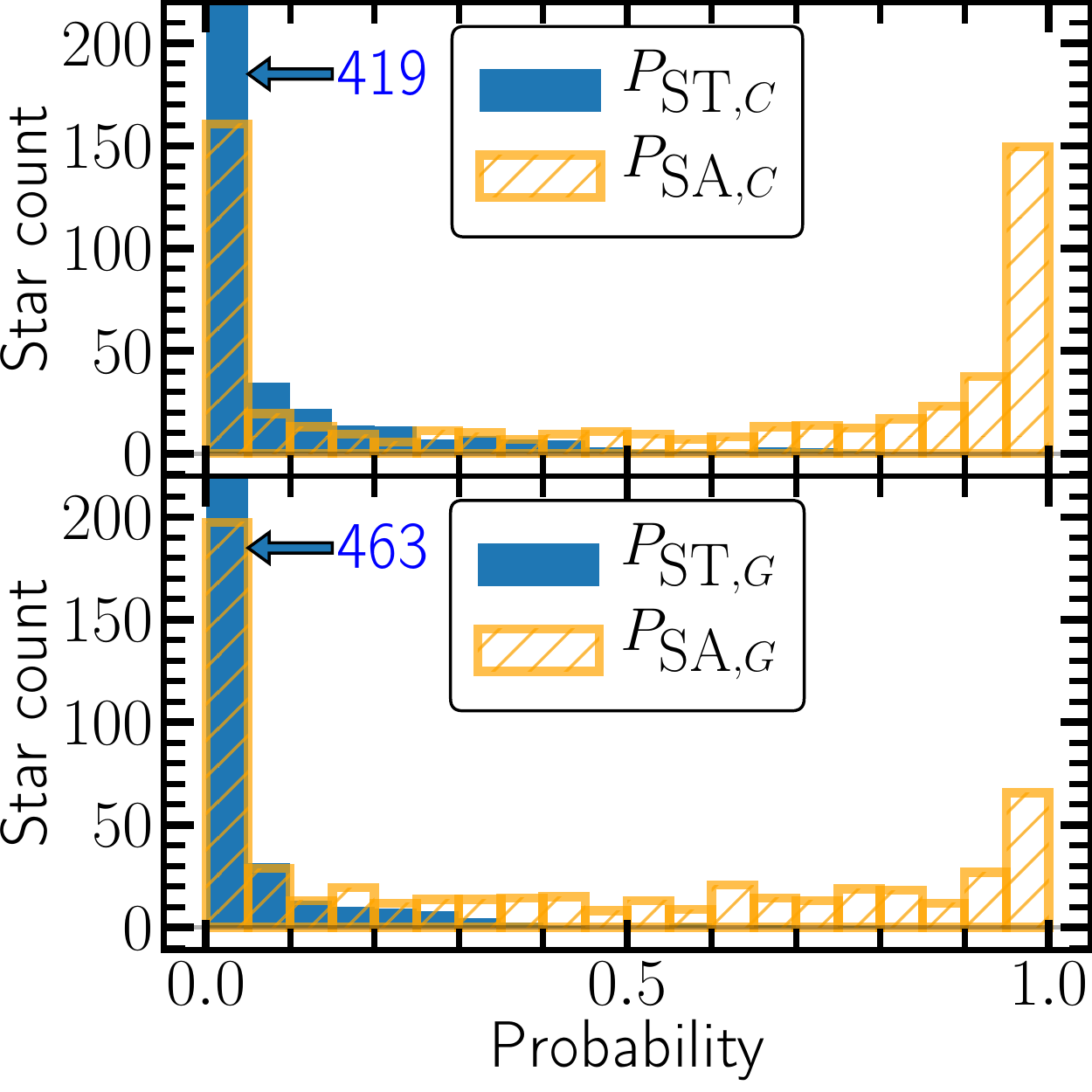}
    \caption{Probability of stars to be solar twins (ST) and solar analogues (SA) within the faint sample of 547 candidate stars. The upper panel shows the probabilities using the spectroscopic calibration \citep[][]{Casali2020} while the lower panel uses the photometric calibration \citep[GALAH DR2,][]{Buder2018}. 
    The probability $P_\textrm{ST} = 0\%-5\%$ bin is only partially shown and peaks at 419 and 463 stars respectively. Solar analogues have about a 30\% rate (in the upper panel) to be in the $P_\textrm{SA} = 95\%-100\%$ bin, which is in agreement with expectations from the photometric selection (SDST2).}
    \label{fig:SP13}
\end{figure}

We identify solar twins and analogues by using stellar parameters and their associated uncertainties to calculate the probability for each star to be a solar twin and analogue. We assume Gaussian probability functions defined by stellar parameter uncertainties, e.g.\ the probability of $\Teff$ values to be within the solar twin definition is:
\begin{align}
    P(\Teffnom\pm\SI{100}{\kelvin}) = \frac{1}{\sqrt{2\pi}\sigma_{T\textrm{eff}}}\int_{\SI{5672}{\kelvin}}^{\SI{5872}{\kelvin}} \exp\left[-\frac{1}{2}\left(\frac{x - \Teff}{\sigma_{T\textrm{eff}}}\right)^2\right] dx,
\end{align}
where $\sigma_{T\textrm{eff}}$ is the uncertainty of $\Teff$. When applying this function to \Teff, $\log g$, and [Fe/H] and multiplying their probabilities with each other, we measure the overall probability of a star to match our solar twin (ST) and solar analogue (SA) criteria:
\begin{align}
    P_\textrm{ST}= &P(\Teffnom\pm\SI{100}{\kelvin}) \times P(\log{g_\odot} \pm \SI{0.2}{dex}) \times\notag \\ &P(\mathrm{[Fe/H]_\odot}\pm \SI{0.1}{dex}), \label{eq:ST_probability} \\
    P_\textrm{SA}= &P(\Teffnom\pm\SI{300}{\kelvin}) \times P(\log{g_\odot} \pm \SI{0.4}{dex}) \times\notag \\ &P(\mathrm{[Fe/H]_\odot}\pm \SI{0.3}{dex}). \label{eq:SA_probability}
\end{align}
Note that this assumes stellar parameters independent from each other, which is not completely accurate. However, as shown in \Fref{fig:sp_hr}, any correlations between stellar parameters within our solar analogue definition must be weak; indeed, this is by construction to a certain degree because of the relatively small ranges in stellar parameters considered. Therefore, any under- or over-estimate of the probability is likely to be correspondingly small. Nevertheless, our probability estimate for an individual star should be regarded as indicative.
The most likely solar twins \citep[using the spectroscopic calibration, ][]{Casali2020} and their stellar parameters are shown in \Tref{tab:best_ST} and a histogram of all solar twin and analogue probabilities is illustrated in \Fref{fig:SP13}.

The star with the highest probability of being a solar twin is Gaia DR3 6197723157722392576 with $P_\textrm{ST}=0.88$ at a distance of $\SI{1.985}{kpc}$. Furthermore, we find 12 (20 including the bright sample) stars in total that are likely solar twins with $P_\textrm{ST}>0.5$. Note that the solar twin and analogue probabilities only include the main stellar parameters and no further selection criteria, like age, which are discussed in \Sref{sec:age_indicator_measure}. While the SDST stellar parameter uncertainties limit the probability to which we can identify solar twins to $P_\textrm{ST}\leq0.88$, the same is not true for solar analogues, which are calculated to have a likelihoods of $P_\textrm{SA}>99.9999\%$ in some cases. We identified 206 (299 including the bright sample) candidate stars as likely ($P_\textrm{SA}>0.9$) solar analogues. The number of solar analogues found at each $G$ magnitude is also in accordance with our previous predictions with photometric data (SDST2, figure 12), i.e. the success rate for selecting solar analogues is between $30\%$ and $60\%$ depending on the $G$ magnitude.
\begin{table*}\label{tab:best_twins_casali}
    \centering
	\begin{tabular}{cccccccccc} 
		\hline
       	Gaia eDR3 identifier & \Teff & $\log g_C$ & [Fe/H] & Age & $\ALi$ & broadening & G & $P_{\textrm{ST, }C}$ \\
         & [K] & [dex] & [dex] & [Gyrs] & [dex] & [km/s] & [mag] & [\%] \\
       	\hline
       	6197723157722392576 & $5782 \pm 61$ & $ 4.51 \pm 0.04 $ & $ -0.016\pm 0.035 $ & $ 2.5 \pm 1.6 $ & $ 1.48 \pm 0.19 $ & $ 5.30 \pm 0.09 $ & $16.05$ & 88.7 \\
       	6005435208437589632 & $5772 \pm 55$ & $ 4.31 \pm 0.05 $ & $ 0.051 \pm 0.027 $ & $ 8.3 \pm 1.2 $ & $ 1.49 \pm 0.18 $ & $ 5.15 \pm 0.08 $ & $15.58$ & 82.9 \\
       	6197681792894913792 & $5709 \pm 48$ & $ 4.38 \pm 0.04 $ & $ -0.051 \pm 0.026 $ & $ 8.3 \pm 1.9 $ & $ 1.14 \pm 0.18 $ & $ 4.72 \pm 0.06 $ & $15.56$ & 75.9 \\
       	6005517603089518592 & $5784 \pm 62$ & $ 4.31 \pm 0.05 $ & $ -0.050 \pm 0.032 $ & $ 9.1 \pm 1.4 $ & $ 1.24 \pm 0.19 $ & $ 5.41 \pm 0.07 $ & $13.58$ & 75.7 \\
       	6197742437833213568 & $5792 \pm 47$ & $ 4.54 \pm 0.03 $ & $ -0.078 \pm 0.028 $ & $ 1.6 \pm 1.1 $ & $ 1.74 \pm 0.12 $ & $ 5.40 \pm 0.07 $ & $13.68$ & 74.9 \\
       	6197742643991631744 & $5836 \pm 54$ & $ 4.32 \pm 0.04 $ & $ -0.006 \pm 0.028 $ & $ 8.1 \pm 1.1 $ & $ 1.58 \pm 0.14 $ & $ 5.25 \pm 0.06 $ & $16.21$ & 72.7 \\
       	6005408132963970176 & $5714 \pm 56$ & $ 4.31 \pm 0.05 $ & $ -0.026 \pm 0.029 $ & $ 9.7 \pm 1.5 $ & $ 1.64 \pm 0.15 $ & $ 5.19 \pm 0.08 $ & $17.04$ & 71.0 \\
       	6197677223051531392 & $5711 \pm 42$ & $ 4.30 \pm 0.04 $ & $ 0.072 \pm 0.021 $ & $ 9.0 \pm 1.0 $ & $ 1.30 \pm 0.15 $ & $ 4.81 \pm 0.04 $ & $16.22$ & 70.2 \\
       	6005389784863513600 & $5797 \pm 72$ & $ 4.36 \pm 0.06 $ & $ -0.056 \pm 0.039 $ & $ 8.0 \pm 2.0 $ & $ 1.70 \pm 0.15 $ & $ 5.16 \pm 0.08 $ & $15.74$ & 69.6 \\
       	6197849983813131648 & $5751 \pm 79$ & $ 4.55 \pm 0.04 $ & $ -0.040 \pm 0.046 $ & $ 2.5 \pm 1.7 $ & $ 1.51 \pm 0.21 $ & $ 5.10 \pm 0.11 $ & $16.05$ & 69.4 \\
        \hline
	\end{tabular}
	\caption{Best solar twins in the sample of candidates with their stellar parameters and likelihoods to be solar twins. The uncertainties in this table are the 1-$\sigma$ errors for all parameters. Not all available parameters are presented here (e.g.\ $P_\textrm{SA}$ and $P_\textrm{ST, nc}$), but they are present in the full table version. The Gaia eDR3 identifier and $G$ band magnitude are from \citet[][]{Collaboration2021}. $\Teff$, $\log g$, and [Fe/H] are measured with $\EPIC$ (SDST1). The age is measured with q$^2$, the lithium abundance $\ALi$ and the line broadening are measured as described in \Sref{sec:activity_indicator}. The solar twin probability is calculated as described in \Sref{sec:new_ST}. The complete table with values for all 877 SDST targets is provided in the supplementary material (online) and in \citet[][]{CLehmann942022a}.}\label{tab:best_ST}
\end{table*}

\subsection{Verification of measurements}\label{sec:verification}
\begin{figure}
    \centering
    \includegraphics[width=0.445\textwidth]{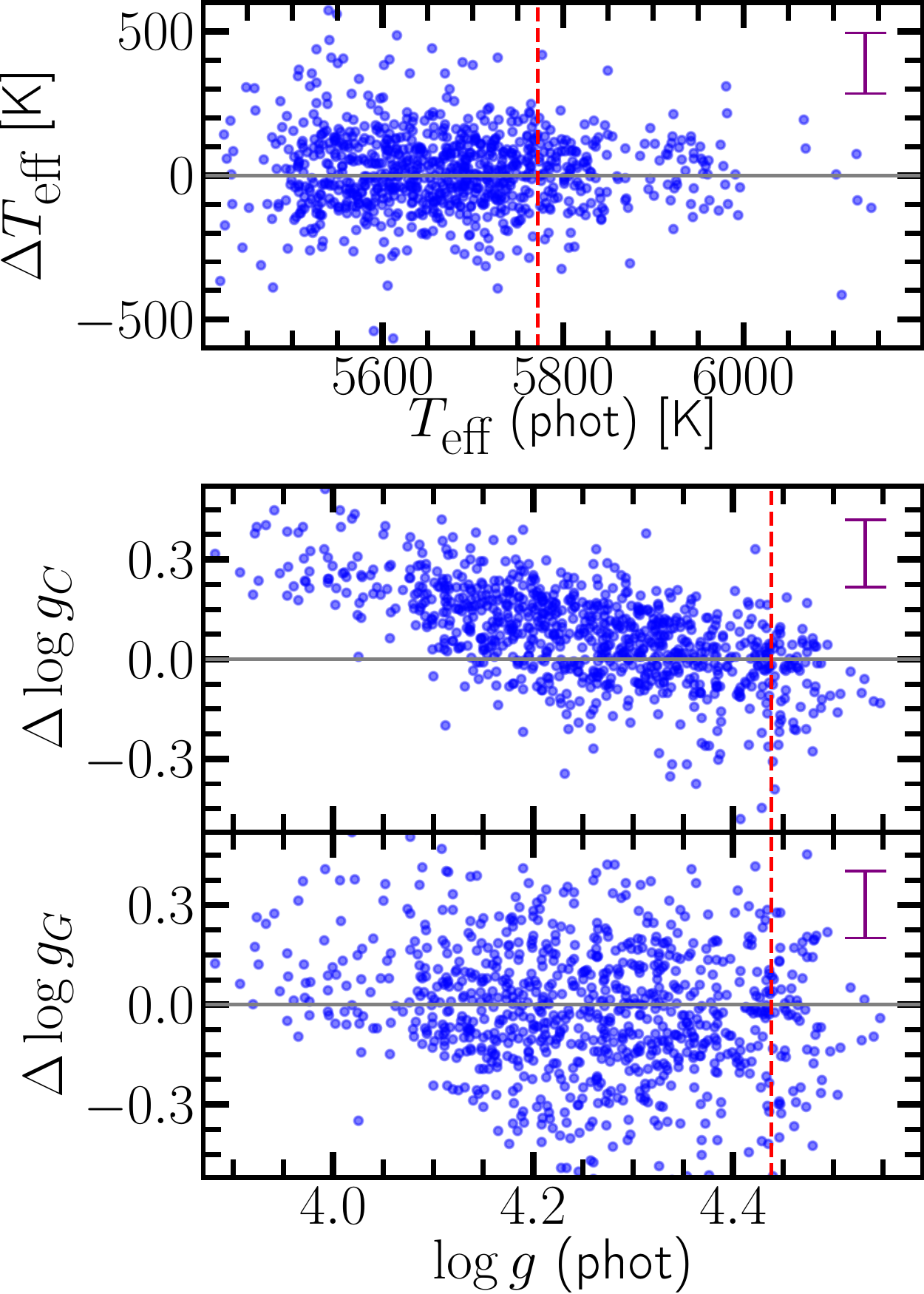}
    \caption{Difference between the photometric and $\EPIC$ effective temperature and surface gravity measurements. The median combined error (photometric and \EPIC\ error) is indicated at the top right of the plot in purple. The top panel shows temperatures, which are in agreement between \EPIC\ and photometric values. The middle and bottom panel show the difference between \EPIC\ and photometric $\log g$ measurements. In the middle panel, \EPIC\ is calibrated on \citet[][]{Casali2020} measurements (i.e.\ spectroscopic surface gravity) while the bottom panel uses the initial GALAH DR2 calibration (i.e.\ photometrically calibrated surface gravity). The red dashed lines indicate solar values.}
    \label{fig:T_logg}
\end{figure}

\subsubsection[Effective temperature]{Effective temperature \Teff}
We verify the SDST effective temperature measurements via a comparison with photometrically determined \Teff\ values \citep[Gaia eDR3,][]{Collaboration2021}. The top panel of \Fref{fig:T_logg} shows the \Teff\ differences between $\EPIC$ measurements and parameters measured with Gaia photometry. Deviations between these measurements -- the RMS and mean error being $145 \pm \SI{105}{\kelvin}$ -- are largely consistent with expectations from the statistical uncertainties alone (combined from the two measurements). There is also no evidence for systematic deviations between the two sets of measurements. Therefore, we conclude from this test that our effective temperature measurements are well calibrated.

\subsubsection[Surface gravity]{Surface gravity $\log g$}
The surface gravity depends considerably on the method of measurement, i.e. photometric or spectroscopic calibration \citep[][]{Adibekyan2012, Bensby2014, Casali2020}. In SDST1 we describe how we calibrate the $\EPIC$ stellar parameters, mainly affecting $\log g$, to match the \citet[][]{Casali2020} measurements, which means that the final $\EPIC$ measurements are calibrated to their spectroscopic method. The initial calibration of $\EPIC$ is based on the combined spectra from \citet[][]{Zwitter2018}, which are made by stacking spectra from GALAH DR2 \citep[][]{Buder2018}. GALAH's measured stellar parameters are calibrated to replicate photometric surface gravities and temperatures to match the main sequence. Therefore, we expect a better match to photometric surface gravities when using the GALAH DR2 calibration of $\EPIC$. 

This is observed in the lower two panels of \Fref{fig:T_logg} which compares both calibrations of $\EPIC$ against the Gaia eDR3 $\log g$ values, and makes the difference in calibration apparent. 
The middle panel reveals an expected slope in the relationship between the (logarithmic) surface gravity derived from photometric measurements (on the x-axis) and that derived from $\EPIC$ using the \citet{Casali2020} spectroscopic measurements as the calibration sample (difference on the y-axis). The discrepancy between photometric and spectroscopic calibrations is a well-known problem \citep[e.g.][]{Bensby2014} which generally increases the more $\log g$ departs from the solar value. In SDST1 we effectively modelled this difference with a linear relationship which is seen directly in the middle panel. In addition, the \citet[][]{Casali2020} sample does not contain stars with low (spectroscopic) $\log g$ values, so the calibration of $\EPIC$ $\log g_C$ values below $\approx$4.24 is an extrapolation.

While we cannot offer a solution to the systematic discrepancy of up to $\SI{0.3}{dex}$ at low $\log g$ between photometric and spectroscopic surface gravities, we acknowledge them here and use the most appropriate calibration for different considerations below. 
The bottom panel shows general agreement between the initial calibration of $\EPIC$ and photometric surface gravity measurements but with large uncertainties, which is not surprising for the reasons mentioned above.

\subsubsection[Metallicity]{Metallicity \textrm{[Fe/H]}}\label{sec:metal_simulation}
\begin{figure}
    \centering
    \includegraphics[width=0.445\textwidth]{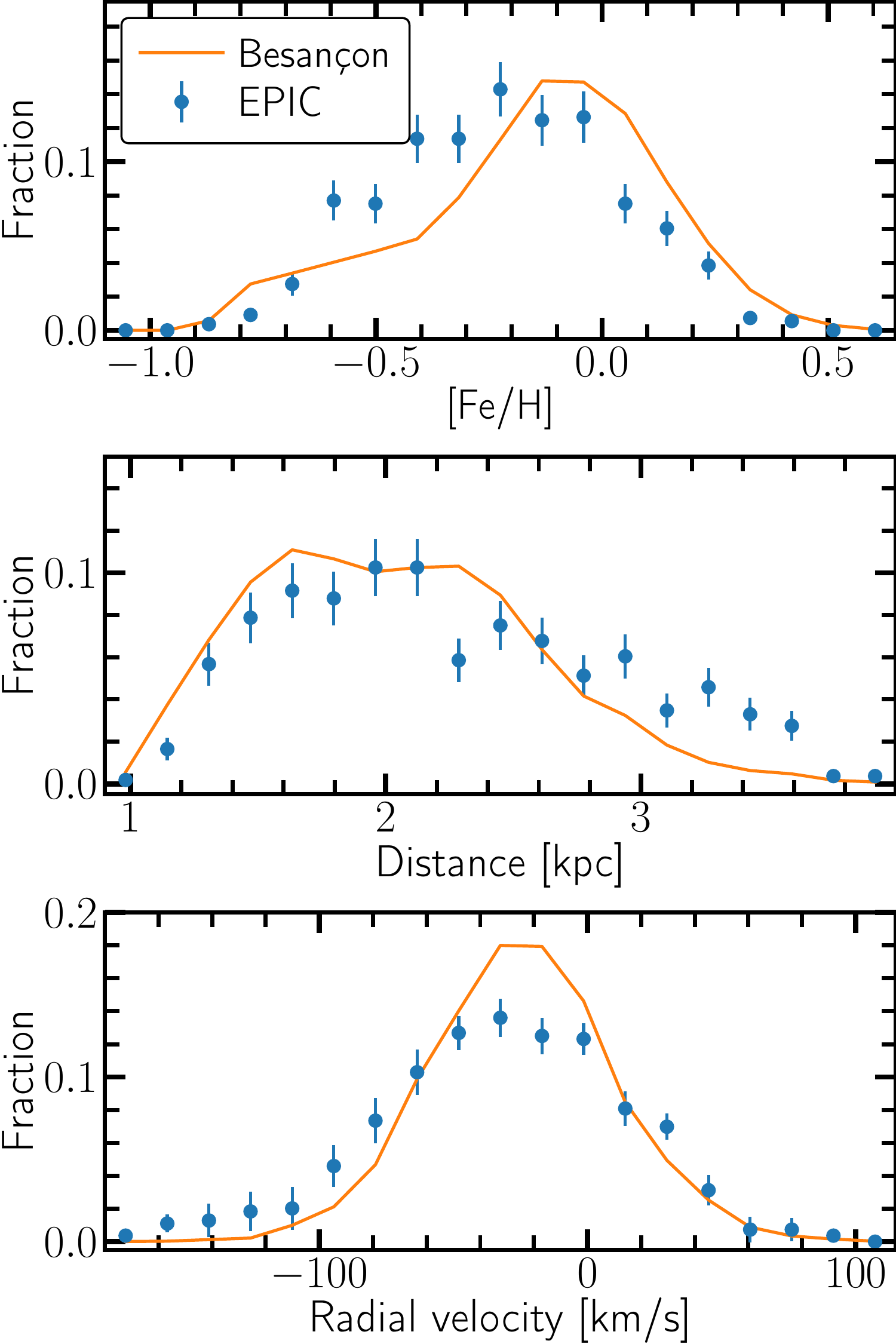}
    \caption{Comparison of metallicity, distance, and radial velocity distribution between the 547 SDST stars in our faint sample (points with error bars) and $9{,}537$ simulated stars from the \besancon\ model (solid curve) in the same target field after applying the transformed selection criteria from SDST2 (see \Sref{sec:metal_simulation}). The errors are statistical uncertainties from the limited number of stars in our sample, i.e.\ $\sqrt{\textrm{number of stars}}$. Upper panel: metallicity. Middle panel: distances; for SDST they are the Gaia measured distances using parallax and photometric values ($r_\textrm{photogeo}$). Lower panel: radial velocity, measured by \EPIC\ for SDST.}
    \label{fig:fe_h_besancon}
\end{figure}
Aside from our measurements with $\EPIC$, there are no reliable measurements of metallicity for the observed stars, so a direct verification of the \EPIC\ measurements is not possible. Instead, we compare the measured SDST metallicities to simulated distributions from the \besancon\ model of stellar population synthesis \citep[][]{Bienayme2018}. The model includes populations of the thin and thick disk as well as halo stars in the same field where we have conducted observations. In addition to the metallicity distribution, it is also instructive to compare the distance and radial velocity distributions of our sample to the \besancon\ model. The same selection criteria (SDST2) used to select SDST targets are applied to the \besancon\ model outputs to allow this statistical comparison. These criteria are based on Gaia photometry and astrometry as well as SkyMapper photometry. The \besancon\ model currently only outputs optical photometric information in the Johnson system, so we transformed the selection criteria in SDST2 to this system using the Gaia EDR3 documentation \citep[][]{Collaboration2021}. We found no valid transformation between the Johnson system and the dereddened SkyMapper $(v-g)_0$ colours, so we constructed $(v-g)_0$ using its dependency on metallicity and temperature \citep[][]{Casagrande2019a}, which are known in the \besancon\ model. The dereddened SkyMapper $(g-z)_0$ criterion in SDST2 was ignored because there is no valid transformation to the Johnson system available. The $(g-z)_0$ selection was meant to eliminate a small number of stars, so ignoring this criterion should not affect the comparison with \EPIC\ results systematically.

For our most distant targets ($\ga\SI{2.5}{kpc}$), the uncertainties in the Gaia parallax measurements become substantial (up to $\sim$50\%; SDST2) so it is important to apply realistic scatter to the \besancon\ model outputs to provide a more accurate comparison with our selected targets. Following the approach detailed in SDST2, the uncertainties in the properties of a model star are taken from a randomly selected real star in the Gaia and SkyMapper surveys (in our target field). These uncertainties were used to add Gaussian-random deviations to the distance, magnitude and colour outputs from the \besancon\ model. Similarly, we used the measured stellar parameter uncertainties from \EPIC\ to add realistic scatter to the stellar parameters of each star in the model. Finally, 10 realisations of the \besancon\ model ($9{,}537$ selected, simulated stars) were combined to reduce the Poisson noise in the results to a negligible level.

The model results are compared with the metallicity, distance and radial velocity distributions of our sample in \Fref{fig:fe_h_besancon}. In general, there is agreement between the expected distributions, based on the model, and the observations. Nevertheless, the distribution of [Fe/H] for our SDST targets (derived from \EPIC) is skewed to slightly lower metallicities, on average, compared to the model. The explanation for this discrepancy is most likely due to an additional, unmodelled observational bias in our SDST sample: we assigned priority to distant targets for observations with HERMES (they were deliberately observed in all exposures to increase their S/N as much as possible) and had to abandon a configuration of brighter stars due to weather and technical problems, which lead to a higher relative number of distant ($\ga\SI{2.5}{kpc}$) targets compared to the \besancon\ model.
This bias is evident in the distance distribution in \Fref{fig:fe_h_besancon} (middle panel). This, in turn, will cause a larger proportion of thick-disk stars to be amongst our targets, as the line of sight to our target field rises out of the thin disk at larger distances. The thick-disk stars will have a lower metallicity, on average, than the thin-disk stars, which explains the skew to lower metallicities in the SDST sample in \Fref{fig:fe_h_besancon} (top panel). Furthermore, this should lead to a broader radial velocity distribution; evidence of this effect is seen in the lower panel of \Fref{fig:fe_h_besancon}. As a higher proportion of SDST stars have ${\rm [Fe/H]} < -0.3$ than expected, the number of solar analogues identified in our survey is reduced. In SDST2 we showed that if this effect is taken into account, the predicted success rate in identifying solar analogues aligns closely with the observed one. Nevertheless, we also note that, given the very specific nature of the SDST targets (both in Galactic location and photometric selection criteria), it is not clear how accurately we should expect the \besancon\ model distributions to match observations. It therefore remains possible that some of the discrepancy in the SDST and \besancon\ model [Fe/H] distributions arises from uncertainties in the latter.

\subsubsection{Stellar parameter uncertainties}\label{sec_SP_uncertainties}
\begin{figure}
    \centering
    \includegraphics[width=0.445\textwidth]{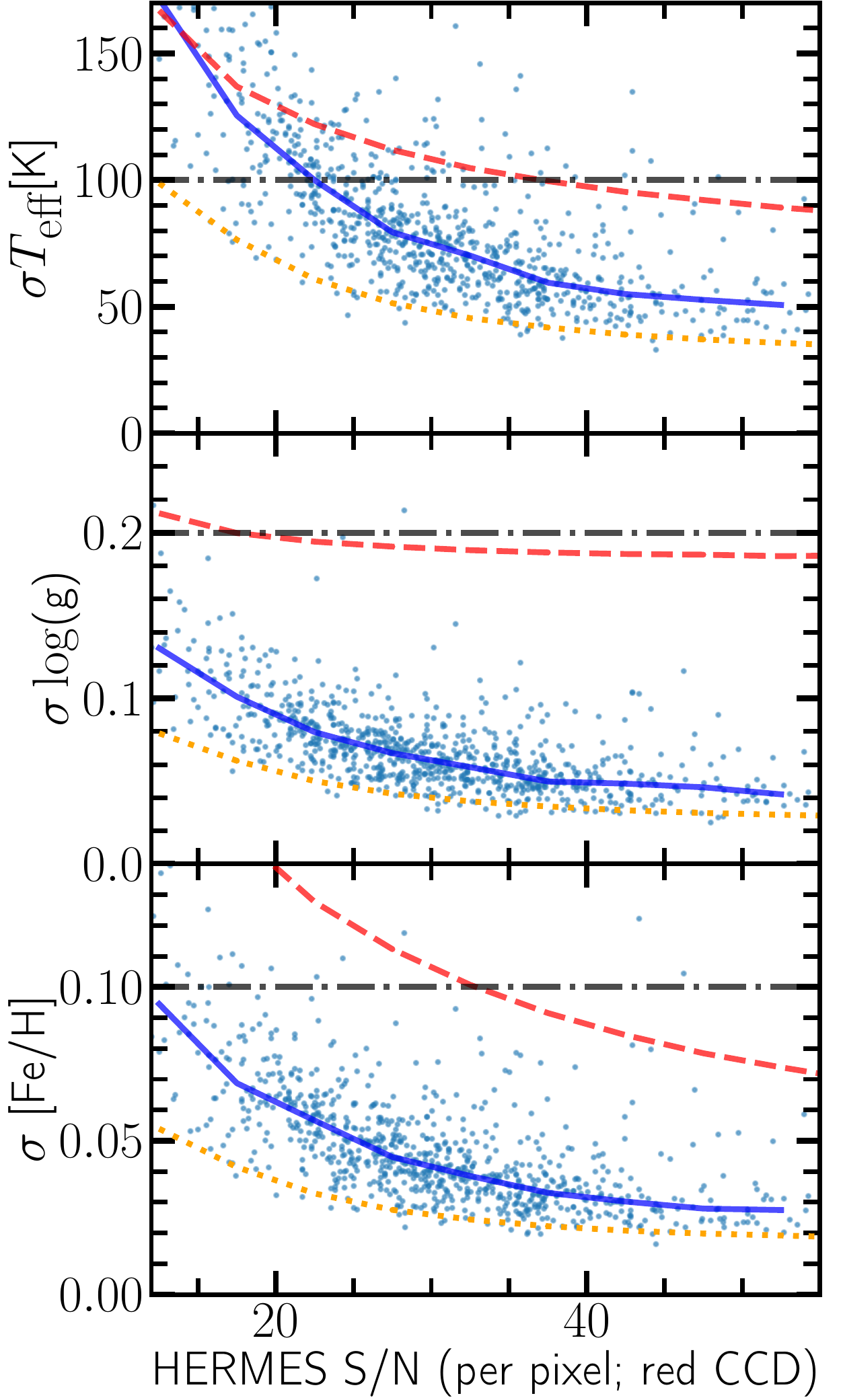}
    \caption{Statistical uncertainties ($\sigma$) for stellar parameters measured with the \EPIC\ algorithm as a function of $S/N$ in the red HERMES band. The blue data points represent the uncertainties \EPIC\ calculated for the SDST set of 877 HERMES spectra while the blue solid line is the median of each $S/N$ bin (bin size=10). The orange dotted line represents the median uncertainty of \EPIC\ when applied to GALAH spectra. The red dashed line shows the median uncertainty for GALAH/The Cannon stellar parameters \citep[][]{cannon2015, Buder2018}. The black dash-dotted horizontal line represents our initial uncertainty target for $S/N\gtrsim25$.}
    \label{fig:uncertainties}
\end{figure}
\begin{figure}
    \centering
    \includegraphics[width=0.445\textwidth]{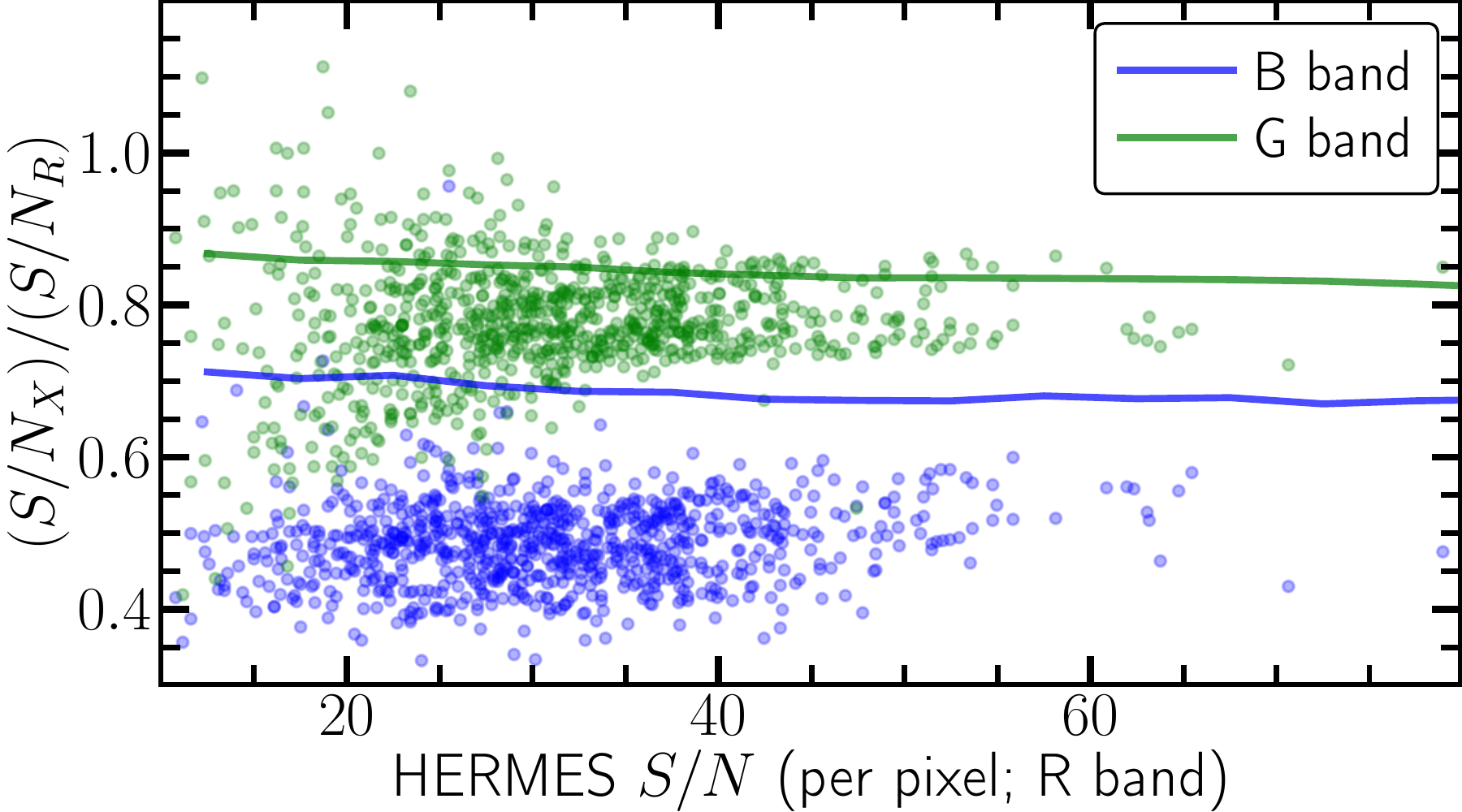}
    \caption{The ratio of the $S/N$ in the $B$ (blue dots) and $G$ (green dots) bands versus their $R$ band $S/N$ for the 877 SDST stars as a function of their $R$ band $S/N$. The lines represent the median relative $S/N$ for solar analogues [according to \Eref{eq:SA_def}] in the GALAH survey.}
    \label{fig:SNR_bands}
\end{figure}
The \EPIC\ uncertainties for the three main stellar parameters are presented in \Fref{fig:uncertainties} (blue points), with their running median shown as the blue solid line. The red dashed lines represent the median uncertainties of GALAH DR3, which are higher than our own measurements with GALAH DR3 spectra (SDST1, orange line) and those for SDST spectra. We initially set the project goal for these uncertainties to $\sigma\left(\Teff, \log g, \textrm{[Fe/H]}\right) \approx \left(\SI{100}{\kelvin, \SI{0.2}{dex}, \SI{0.1}{dex}}\right)$ at $S/N=25$ in the $R$ band so that we can identify solar analogue stars with 2--$3\sigma$ confidence in each stellar parameter. The black dash-dotted horizontal line in each panel indicates this goal. It is clear that our measurements meet the goals at all $S/N$, though the effective temperature uncertainties at $S/N\lesssim22$ are above the $\SI{100}{\kelvin}$ level.

The SDST uncertainties are larger than the initial expectation from SDST1 (orange line in \Fref{fig:uncertainties}), especially for \Teff. The reason is that, for a given $R$ band $S/N$, the $B$ and $G$ bands within SDST observations have lower $S/N$ compared to spectra from GALAH DR3. This is demonstrated in \Fref{fig:SNR_bands}, which shows the measured $S/N$ for the 877 stars in SDST and the median $S/N$ for solar analogue stars in GALAH DR3 \citep[][]{Buder2021}. 
The $G$ band $S/N$ is typically $\approx80\%$ of the $R$ band $S/N$ for SDST targets, compared to $\approx85\%$ in GALAH. The difference is larger for the $B$ band: $\approx50\%$ for SDST versus $\approx70\%$ for GALAH. As discussed in SDST2, this degradation of the $S/N$, particularly in these two bands, is due to the larger relative contribution of CCD read noise and scattered sky light for the much fainter targets of SDST (see \Fref{fig:Gmag_hist} for a comparison of the SDST and GALAH magnitude distributions). Indeed, \Fref{fig:SNR_bands} also shows the expected decline in the $B$ band $S/N$ (relative to the $R$ band) towards lower $R$ band $S/N$, i.e.\ for the faintest targets. However, this effect is reduced to some extent by our strategy to observe fainter targets in a larger number of exposures.

\subsection{Distance biases}
\begin{figure}
    \centering
    \includegraphics[width=0.445\textwidth]{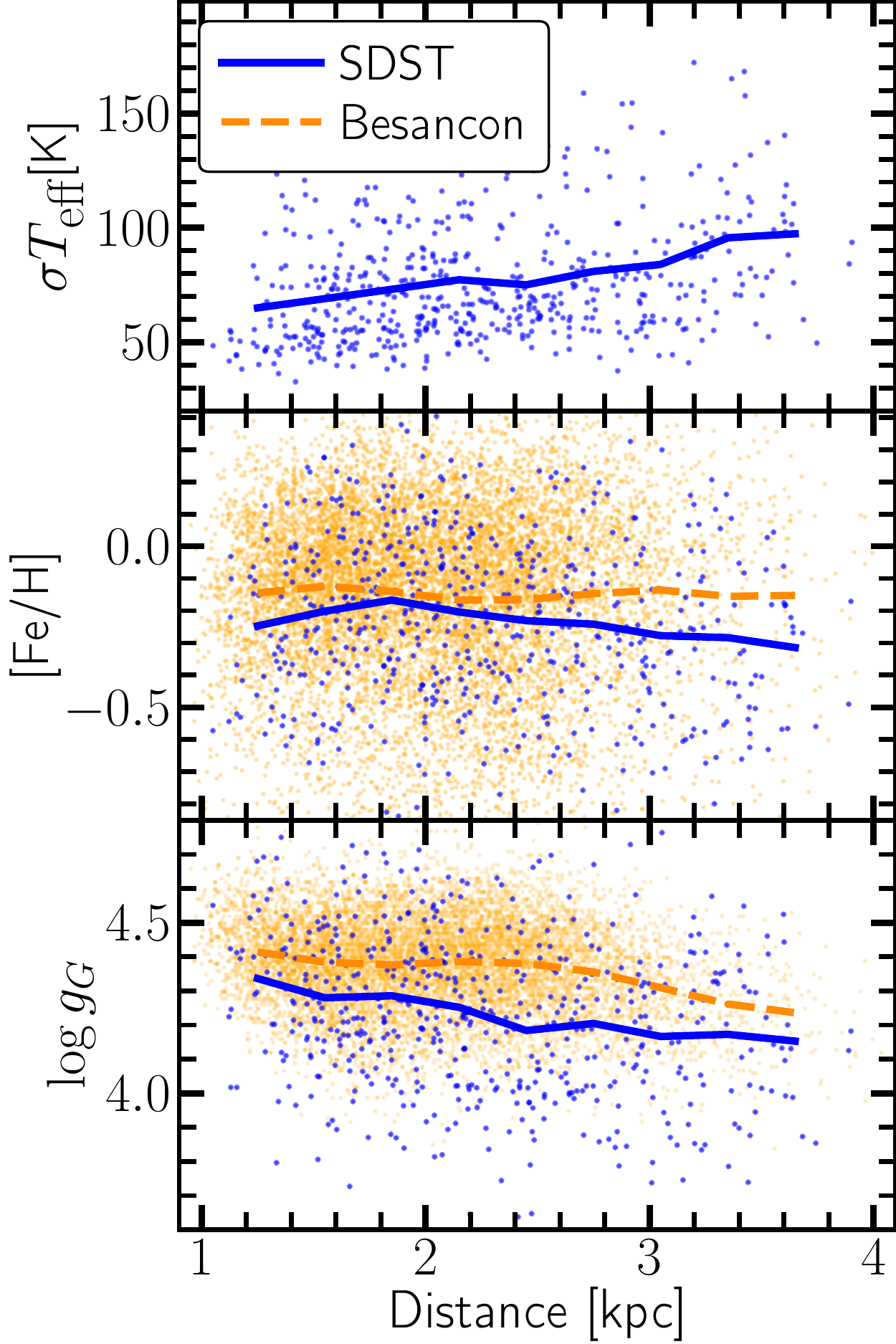}
    \caption{Stellar parameters and uncertainties as a function of the distance. Each panel shows $\EPIC$ measurements (blue dots) and the middle/bottom panels also show the \besancon\ model simulation results (orange dots). The solid and dashed lines show a binned average (distance bins of $\SI{500}{pc}$). Top panel: the temperature uncertainty; middle panel: the metallicity; bottom panel: the surface gravity using the initial (photometric) calibration of \EPIC\ with GALAH DR2.}
    \label{fig:distance}
\end{figure}
In this section, we consider biases in our stellar parameter measurements that arise because of the large distances probed by SDST. While highlighting these main biases in our current results, this will also be informative for future surveys for distant solar twins and analogues. It is important to keep in mind that the target field (centred on RA $227.1^\circ$, Dec $-39.0^\circ$) is off the Galactic Plane and therefore contains both thin and thick disk stars especially at the largest distances probed (see \Sref{sec:metal_simulation} and \Fref{fig:fe_h_besancon}).

Firstly, our stellar parameter uncertainties increase with distance, since the target stars are all nearly the same absolute magnitude. This holds true, even though additional observation time was spent to increase the $S/N$ for distant targets, because of the increased relative influence of read, dark, and sky noise in the spectra (the same effect that caused larger uncertainties compared to GALAH spectra in \Fref{fig:uncertainties} and \ref{fig:SNR_bands}). As a direct result, there is less confidence in a solar twin and analogue identification which is reflected in lower twin and analogue probabilities [\Eref{eq:ST_probability} and (\ref{eq:SA_probability})]. As an example, the top panel of \Fref{fig:distance} shows the temperature uncertainty as a function of distance with an average uncertainty increase from $\sim\SI{60}{\kelvin}$ to $\sim\SI{100}{\kelvin}$ over our distance range.

Parameters like stellar age, metallicity and radial velocity also change with distance due to a shift in the stellar population sampled. At larger distances, our field probes stars that are 1--$\SI{1.5}{kpc}$ off the Galactic disk, which leads to a larger fraction of observed stars being thick disk stars, i.e.\ $\sim42\%$ of distant stars (distances 3--$\SI{4}{kpc}$) are in the thick disk compared to only $\sim25\%$ of nearby stars (distances 1--$\SI{2}{kpc}$) according to the \besancon\ model. Note that the measurements of \Sref{sec:verification} suggest that the fraction of thick disk stars may be higher than these predictions.
Thick disk stars tend to be older, more metal poor and have a larger dispersion in radial velocity, therefore we expect the distribution of these parameters to change as a function of distance. Nevertheless, the metallicity of SDST targets shows only a marginal and statistically insignificant decrease, on average, at larger distances, as shown in \Fref{fig:distance} (middle panel).

Lastly, parameters like effective temperature and surface gravity influence the luminosity of stars. Our photometric and astrometric target selection criteria include apparent and absolute magnitude cuts, which means that the most distant stars will tend to be brighter (SDST2) and, on average, hotter and larger, i.e.\ lower surface gravities. This effect is shown in the bottom panel of \Fref{fig:distance} where the average surface gravity drops from $\log g\approx\SI{4.3}{dex}$ for nearby stars (1--$\SI{2}{kpc}$) down to $\log g \approx 4.1$ at larger distances (3--$\SI{4}{kpc}$). This means there is a significant reduction in the number of stars with solar surface gravities in our sample at these large distances.

The above distance biases are not a concern for measurements of $\alpha$: \citet[][]{Berke2022b} has shown that $\alpha$ can be measured reliably for stellar parameters within the solar analogue classification, \Eref{eq:SA_def}. However, these biases could be minimised in future selections for solar twin surveys. For example, the selection process in SDST2 could be adjusted to use distances of stars rather than their Gaia $G$ magnitude for the selection to reduce the bias towards brighter targets at large distances. 
Note that this would also increase the observation time considerably when aiming to acquire similar $S/N$ spectra at all distances.
Furthermore, it is possible to minimise the effects of thick disk stars on our sample if stellar velocities are available for potential targets, as thin and thick disk stars are kinematically distinct to a certain degree.

\subsection{Indicators for fast rotating and active stars}\label{sec:age_indicator_measure}
\begin{figure}
    \centering
    \includegraphics[width=0.49\textwidth]{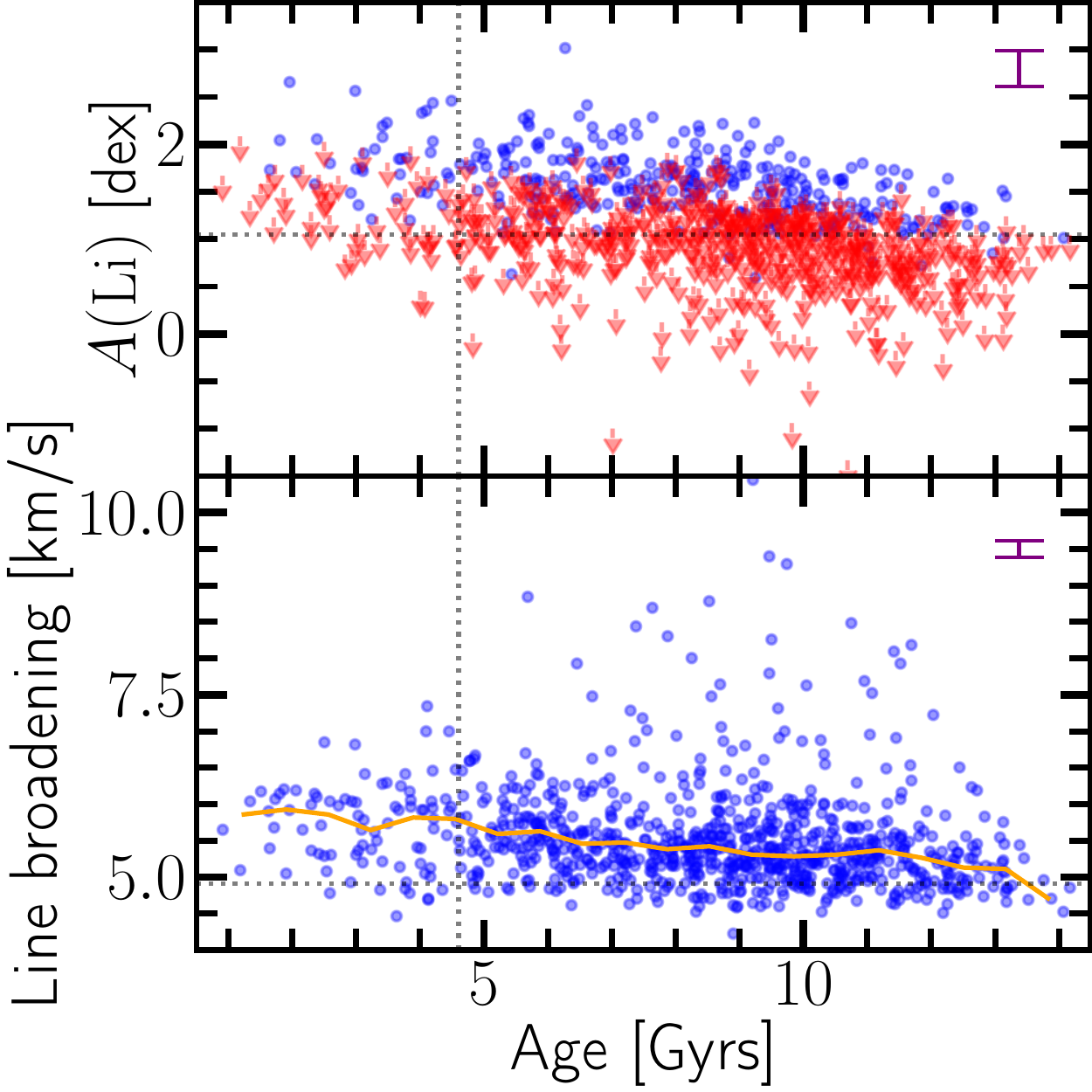}
    \caption{Age indicators, measured as described in \Sref{sec:activity_indicator}, as a function of their age derived using the q$^2$ algorithm. The top panel shows lithium abundances $\ALi$ while the bottom panel shows line broadening. Most lithium measurements are only upper limits, represented by red arrows (non-upper limit measurements are blue dots). The orange solid line in the bottom panel shows a running median for the broadening. The dotted black lines show the median solar measurement for the line broadening ($b=\SI{4.9}{\kilo\metre\per\second}$) as well as literature values for solar lithium abundance [$\ALi=\SI{1.05}{dex}$, \citealt[][]{Lodders2019}] and age \citep[$\textrm{age}=\SI{4.6}{Gyrs}$, ][]{Connelly2012}. The median error (of non-upper limit measurements) is indicated in the top right of each panel.}
    \label{fig:age_plot}
\end{figure}
\begin{figure}
    \centering
    \includegraphics[width=0.49\textwidth]{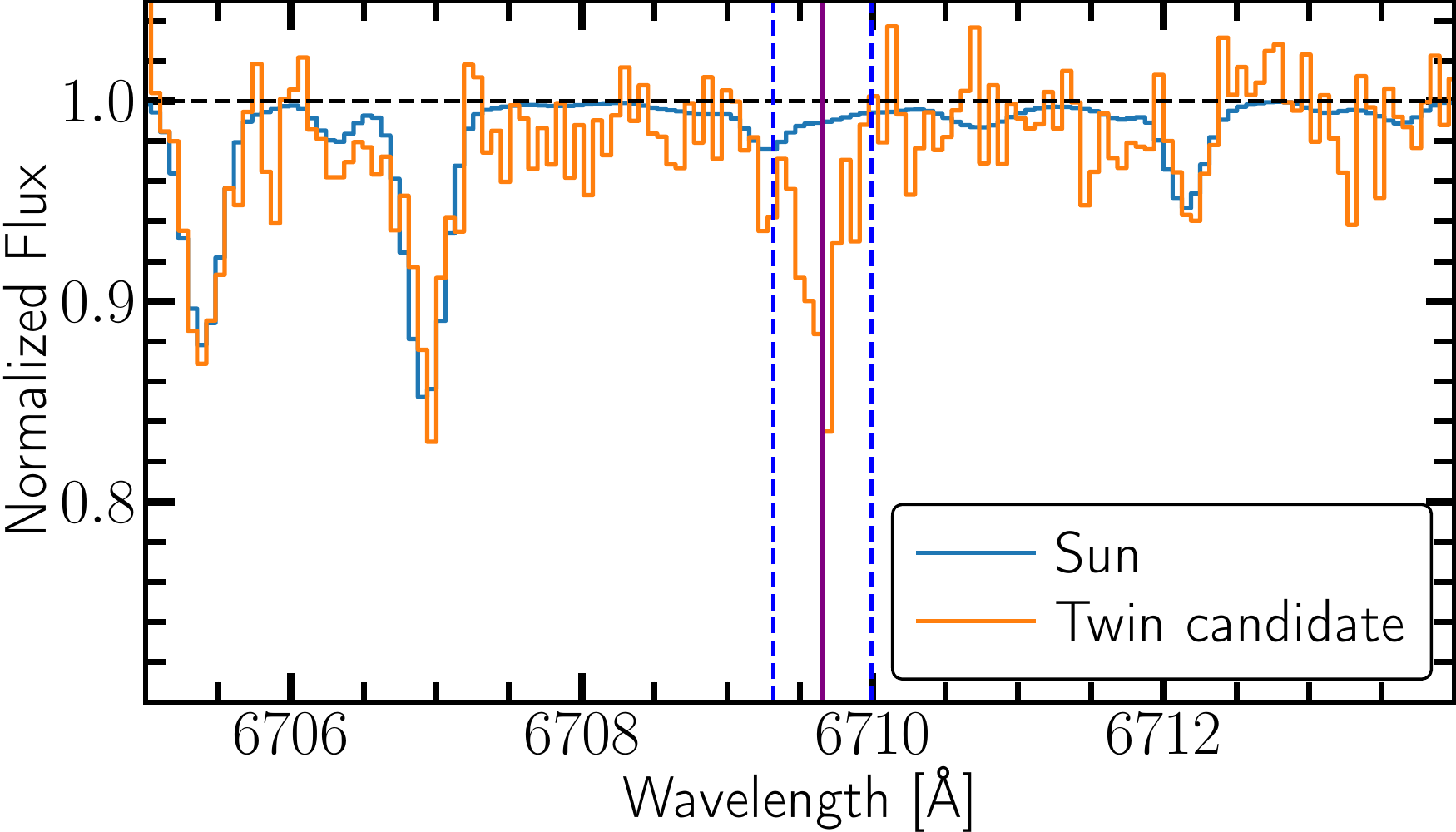}
    \caption{Apparent, strong lithium detection in a SDST star (orange spectrum) compared to the solar reference spectrum \citep[blue spectrum, KPNO2010 in][]{Chance2010}. The purple vertical line shows the expected centre of the line and the blue dashed lines show the area within which $\EPIC$ measures the EW (SDST1). This example lithium feature has a measured $\ALi=\SI{2.16}{dex}$ and, for comparison, the solar reference spectrum [$\ALi=\SI{1.05}{dex}$] has a lithium feature that is not detectable in low $S/N$ spectra.}
    \label{fig:Li_example}
\end{figure}
The determination of fast rotation, activity, and spectroscopic binarity of stars is important because active stars tend to have broader, variable spectral line profiles, which is not ideal for a precise spectroscopic measurement of $\alpha$ \citep[][]{Berke2022b, Berke2022a}.
We can select against active, faster-rotating or spectroscopic binary stars using their measured spectral line broadening and stellar age, because young stars tend to be active and have fast rotation, slowing down over the duration of their lifetime. Additionally, the lithium abundance is a useful measurement as it correlates strongly with stellar age and provides an additional estimator of stellar youth.
As described in \Sref{sec:activity_indicator}, we measure the lithium abundance, the spectral line broadening, and the ages of our target stars. This gives us the opportunity to verify how well our lithium abundance and broadening measurements correlate with age.

In \Fref{fig:age_plot} we show the lithium abundance and line broadening against the stellar age measurement of the SDST stars. We found that the lithium abundance measurements should only be regarded as upper limits for $\ALi < 1.2$ as the single lithium absorption feature becomes too weak to make a reliable measurement below that threshold (depending on the $S/N$ of the spectrum). In \Fref{fig:Li_example} we show an example for a strong lithium feature in our sample [$\ALi=\SI{2.16}{dex}$] compared to the lithium feature in the solar spectrum [$\ALi=\SI{1.05}{dex}$] of the same resolution as the SDST HERMES spectra. It is clear that a lithium abundance of $\ALi=\SI{1.05}{dex}$ would not be detectable with even a moderate amount of noise in the spectrum, so only upper limit measurements for lithium are possible in most targets. 
As shown in \Fref{fig:age_plot}, most SDST spectra only provide an upper limit on $\ALi$, so it is not possible to assess whetever the expected decline in $\ALi$ with age is observed in SDST stars. Nevertheless, a mild decrease in $\ALi$ with age is present, indicating that our $\ALi$ measurements are reliable enough for the purpose of rejecting the highest $\ALi$ stars when selecting targets for future high-resolution spectroscopy to measure the fine-structure constant (see below).

The lower panel of \Fref{fig:age_plot} shows the line-broadening measurements for the SDST targets. Most stars (82\%) have a measured broadening $<\SI{6}{\kilo\metre\per\second}$ and a mildly decreasing trend with age is visible. However, with the HERMES resolving power ($R\lesssim32{,}000$) corresponding to a ${\rm FWHM} \gtrsim \SI{9.5}{\kilo\metre\per\second}$, our line-fitting approach is unable to reliably measure broadening values $\lesssim\SI{4}{\kilo\metre\per\second}$. Therefore, it is unlikely to reliably reproduce the correct gradient of the decline in broadening with age. Nevertheless, \Fref{fig:age_plot} shows a noticeable group of stars with significantly higher broadening values, $\gtrsim\SI{7}{\kilo\metre\per\second}$. These are likely to be faster-rotating -- and therefore likely younger and more active -- or spectroscopic binary stars. The latter interpretation may be more likely because most of these stars -- and those with the larger broadening values -- have q$^2$ ages $\gtrsim$6\,Gyr. Just like the lithium abundance analysis above, the spectral line broadening is more useful for identifying these stars which should be avoided in follow-up high-resolution spectroscopy for measuring the fine-structure constant.

For each of the SDST stars, \Tref{tab:best_ST} includes our measurements of age (from q$^2$), \ALi\ and line broadening (derived within \EPIC). These can be used as additional selection criteria for identifying the best targets for follow-up measurements of the fine-structure constant. These measurements require differential comparison of line centroids between very similar stars, and have so-far been demonstrated with mostly older stars, $>\SI{2}{Gyrs}$, which show very similar line widths as the Sun, i.e.\ broadening $<\SI{6}{\kilo\metre\per\second}$ \citep[][]{Berke2022b, Berke2022a, Murphy2022}. The latter criterion will also tend to reject some spectroscopic binaries, which are not useful for constraining variations in the fine-structure constant. Note that the expected anti-correlation between stellar rotation speed and age \citep[e.g.][]{Meibom2015} means that rejecting stars younger than $\SI{2}{Gyrs}$ will assist in ensuring no faster-rotating stars are followed up with high-resolution spectroscopy. Stars with $\ALi > \SI{2.0}{dex}$ are also expected to be younger than $\sim\SI{2}{Gyrs}$ \citep[e.g.][]{Carlos2019}, and can be confidently identified in the HERMES spectra (see \Fref{fig:age_plot}). Amongst the 206 SDST stars in the faint sample ($G > \SI{15.4}{mag}$) which are most likely to be solar analogues ($P_\textrm{SA,C}>90\%$), we find that 174 stars satisfy these additional selection criteria [age $>\SI{2}{Gyrs}$, line broadening $<\SI{6}{\kilo\metre\per\second}$, and $\ALi < \SI{2.0}{dex}$].

\section{Conclusions}\label{sec:conclusion}
We have presented the results of our Survey for Distant Solar Twins (SDST), which includes the identification of the most likely solar twins and analogues from the set of 877 candidates. 
For the faint set, we identified 206 out of 547 stars with a high likelihood of being solar analogues ($P_\textrm{SA}>90\%$), with 12 of these targets being solar twin candidates with $P_\textrm{ST}>50\%$. For the bright sample, we found 93 out of 330 likely solar analogues with 8 solar twin candidates.

We found that our measured metallicity distribution is more metal poor than the \besancon\ model predictions, which can be explained with a higher than predicted ratio of thick-to-thin disk stars in the observations. This is consistent with the observed differences in both the distance and radial velocity distributions between our sample and the model data-set. We conclude that the metallicity distribution of the SDST targets is broadly consistent with expectations, which supports our identification of the new solar twins and analogues.

We selected the lithium abundance, the spectral line broadening and the age of stars to identify possibly very active/fast rotating stars. While none of these three measurements strongly constrain fast rotation and stellar activity individually, together they can be used to select against active stars and spectroscopic binaries for follow-up observations. Applying additional selection criteria -- $\ALi < \SI{2.0}{dex}, \textrm{ broadening} <\SI{6}{\kilo\metre\per\second}\textrm{ and age}>\SI{2}{Gyrs}$ -- to the SDST sample reduces the number of distant solar analogue targets from 206 to 174 (the bright sample solar analogues drop from 93 to 73 and solar twins from 20 to 15 accordingly).

We identified the main biases in the SDST sample as a function of distance. The first is an apparent increase in the ratio of thick-to-thin disk stars at distances $\geq \SI{2.5}{kpc}$. This is expected because the target field lies $16^\circ$ off the Galactic Plane. The second main bias is that the most distant stars have brighter absolute $G$ magnitudes, on average.
This produces a bias in the surface gravity, measured in $\EPIC$, with $\log g$ dropping by $\SI{0.3}{dex}$ on average for the most distant stars ($\sim\SI{4}{kpc}$). This is a selection effect that is hard to counteract as gathering high-enough S/N spectra of stars with solar surface gravities becomes more challenging and time intensive at large distances.

We have also shown that the \EPIC\ method described in SDST1 for measuring stellar parameters can be extended to measure element abundances, i.e.\ the lithium abundance (\Sref{sec:lithium_abundance}). However, lithium has only one absorption feature within HERMES's bands, and that line is not detectable at lower abundances. Future work on measuring element abundances with more lines can potentially increase the precision of abundance measurements, especially if they have more absorption lines in the spectral data.

The new set of stars presented in this work provides opportunities to study solar twins and analogues in the inner Galaxy, not just to probe the fine-structure constant.
While HERMES observations do not have the resolving power and wavelength coverage needed for a detailed analysis of these targets, follow-up observations of the solar analogues identified here, with wider wavelength coverage and higher resolving power, would enable this. 
Such spectra would allow studies of the chemical composition of these stars \citep{Nissen2015, Nissen2016, Nissen2020, Carlos2016, Liu2016, Adibekyan2017}, stellar yields and nucleosynthetic processes \citep[e.g.][]{Karakas2016}, the (non-)universality of chemical clocks and
Galactic chemical evolution across the Milky Way \citep{Kobayashi2006, Casali2020, Vazquez2022}.

\section*{Acknowledgements}
CL, MTM and FL acknowledge the support of the Australian Research Council through \textsl{Future Fellowship} grant FT180100194.

This work is based on data acquired at the Anglo-Australian Telescope under program A/2021A/005. We acknowledge the traditional custodians of the land on which the AAT stands, the Gamilaraay people, and pay our respects to elders past and present.

We acknowledge the use of the following software for the development of the $\EPIC$ algorithm and and any further methods: Python 3.7.4 by the Python Software Foundation \citep[][]{python} as well as the packages Astropy \citep[][]{astropy}, Matplotlib \citep[][]{matplotlib}, Numpy \citep[][]{numpy} and Scipy \citep[][]{scipy}. We have also made use of the TOPCAT software for data visualisation and analysis \citep[][]{Taylor2005}.

\section*{Data Availability}
The SDST catalogue is available at \href{https://doi.org/10.5281/zenodo.7306319}{https://doi.org/10.5281/zenodo.7306319}. The SDST spectra are available in \citet[][]{Liu2022}. The \EPIC\ code is available at \citet[][]{CLehmann942022}.



\bibliographystyle{mnras}
\bibliography{Bibliography} 


\section*{Supporting information}
Supplementary material is available at \href{MNRAS link}{MNRAS} online.
\\\newline
\textbf{SDST\_catalogue\_V1.0.csv} 
\\\newline
Please note: Oxford University Press is not responsible for the content
or functionality of any supporting materials supplied by the authors.
Any queries (other than missing material) should be directed to the
corresponding author for the article.

\appendix


\bsp	
\label{lastpage}
\end{CJK*}
\end{document}